\documentclass[twocolumn]{aastex61}

\shorttitle{Coronal Evolution of AR 11437}
\shortauthors{Simulating the coronal evolution of AR 11437}

\begin{document}

\title{Simulating the coronal evolution of AR 11437 using \textit{SDO}/HMI magnetograms}

\correspondingauthor{Stephanie L. Yardley}
\email{sly3@st-andrews.ac.uk}

\author[0000-0003-2802-4381]{Stephanie L. Yardley}
\affil{School of Mathematics and Statistics, University of St Andrews, North Haugh, St Andrews, Fife, KY16 9SS, UK}

\author{Duncan H. Mackay}
\affiliation{School of Mathematics and Statistics, University of St Andrews, North Haugh, St Andrews, Fife, KY16 9SS, UK}

\author{Lucie M. Green}
\affiliation{Mullard Space Science Laboratory, University College London, Holmbury St. Mary, Dorking, Surrey, RH5 6NT, UK}

\begin{abstract}

The coronal magnetic field evolution of AR 11437 is simulated by applying the magnetofrictional relaxation technique of \citet{Mackay-2011}. A sequence of photospheric line-of-sight magnetograms produced by \textit{SDO}/HMI are used to drive the simulation and continuously evolve the coronal magnetic field of the active region through a series of non-linear force-free equilibria. The simulation is started during the first stages of the active region emergence so that its full evolution from emergence to decay can be simulated. A comparison of the simulation results with \textit{SDO}/AIA observations show that many aspects of the active region's observed coronal evolution are reproduced. In particular, it shows the presence of a flux rope, which forms at the same location as sheared coronal loops in the observations. The observations show that eruptions occur on 2012 March 17 at 05:09 UT and 10:45 UT and on 2012 March 20 at 14:31 UT. The simulation reproduces the first and third eruption, with the simulated flux rope erupting roughly 1 and 10 hours before the observed ejections, respectively. A parameter study is conducted where the boundary and initial conditions are varied along with the physical effects of Ohmic diffusion, hyperdiffusion and an additional injection of helicity. When comparing the simulations, the evolution of the magnetic field, free magnetic energy, relative helicity and flux rope eruption timings do not change significantly. This indicates that the key element in reproducing the coronal evolution of AR 11437 is the use of line-of-sight magnetograms to drive the evolution of the coronal magnetic field. 


\end{abstract}


\section{Introduction} \label{sec:intro}


Coronal mass ejections (CMEs) are the largest eruptive phenomenon in the Solar System, sending $\sim$10$^{12}$ kg of magnetised plasma into interplanetary space at speeds up to a few 1000 km s$^{-1}$. These eruptions are magnetically driven and approximately 10$^{32}$ ergs of free magnetic energy is initially built up in the non-potential coronal magnetic field. At the time of eruption a critical point is reached and equilibrium is lost, resulting in the energy stored being released as a CME \citep{Forbes-2000}.


Currently, all theoretical CME models involve the formation of a flux rope, which is composed of helical magnetic field lines. However, the models differ in when the magnetic flux rope forms. One set of models requires the flux rope to be present prior to the eruption with CMEs being a result of an ideal instability or loss of equilibrium \citep{Forbes-1991, Torok-2005, Kliem-2006, Mackay-2006a, Mackay-2006b, vB-2007}. In the other scenario the flux rope forms in-situ, during the eruption, as a product of magnetic reconnection \citep{Antiochos-1999, Moore-2001}.


There is increasing observational evidence provided by soft X-ray and EUV emission in the corona that flux ropes form prior to the eruption of CMEs \citep{Green-2009, Green-2011, Patsourakos-2013}. The first model of a flux rope was proposed by \citet{Kuperus-1974} and consisted of a filament embedded in a current sheet. This was advanced by \citet{vB-1989} with a model that focuses on how the coronal field evolves in response to the shearing and convergence of photospheric magnetic field. These photospheric motions drive flux cancellation and associated magnetic reconnection at the polarity inversion line (PIL), transforming the sheared magnetic arcade into a flux rope configuration.


As it is currently very difficult to measure the coronal magnetic field directly, an alternative approach, such as simulations or extrapolations of the photospheric magnetic field, must be used to infer the pre-eruptive magnetic structure of CMEs. These numerical methods, which use observational constraints, rely on the corona being approximated as ``force-free". Therefore, the coronal magnetic field satisfies the force-free criterion of $\mathbf{j} \times \mathbf{B} = 0$, where $\mathbf{j} = \alpha \mathbf{B}$. The torsion parameter $\alpha = \alpha (\mathbf{r})$ is a scalar function that remains constant along field lines, but is allowed to vary as a function of position. If $\alpha = 0$, this is the lowest energy case where the magnetic field is potential. While potential fields can provide an approximate description of the coronal magnetic field, they cannot be used to model active regions or filaments as the free magnetic energy needed to drive eruptions is not included. The most realistic description is provided by a non-linear force-free (NLFF) magnetic field where $\alpha = \alpha (\mathbf{r})$ is allowed to vary as a function of position.



A number of NLFF magnetic field methods have been recently developed. The techniques that are used to generate NLFF magnetic fields can be split into two main categories: static or time-dependent models. Static models, which use a single fixed lower boundary condition for the normal field component, either extrapolate the NLFF magnetic field into the corona using vector magnetograms \citep{Regnier-2002, Schrijver-2006, Canou-2010, Jiang-2014} or evolve the initial potential or LFF coronal field into a NLFF state. The latter approach uses the magnetofrictional method \citep{Yang-1986} to produce static NLFF magnetic field models at different snapshots during the evolution of an active region. This can be achieved by taking the potential field extrapolation of a magnetogram, setting the photospheric magnetic field to equal that found in the vector magnetogram \citep{Valori-2005} or by inserting a flux rope into the potential field \citep{Bobra-2008, Su-2009, Savcheva-2012}. In both cases the coronal field is then relaxed to a NLFF state. Although these methods can produce a series of NLFF magnetic field models by changing the lower boundary condition, each model is independent and therefore cannot be used to study the dynamical, quasi-static evolution of the coronal magnetic field with time. 

It is possible to use the magnetofrictional relaxation technique to construct a continuous time-dependent series of NLFF magnetic fields by evolving the initial coronal field through changing the photospheric boundary conditions. The coronal field evolution can either be driven using a continuous time series of artificial \citep{vB-2000, vB-2006, Mackay-2009} or observed \citep{Yeates-2007, Mackay-2011, Gibb-2014} magnetograms. By using this method the memory of the previous magnetic field connectivities can be maintained along with the global conserved quantities. \citet{Gibb-2014} applied the method of \citet{Mackay-2011} to simulate the coronal evolution of AR 10977 using \textit{SOHO}/MDI line-of-sight (LoS) magnetograms as lower boundary conditions and compared the simulated evolution to \textit{Hinode}/XRT observations. They were able to reproduce the main coronal features and time-evolution of the active region up until the single eruption of a sigmoid.  We extend this study by simulating the coronal evolution of a different active region to determine if the simulation can reproduce multiple eruptions that originate from the active region. The timings of the simulated flux rope eruptions and the observed ejections in the \textit{SDO}/AIA data will be compared. We also conduct a parameter study where non-ideal terms and an additional injection of helicity are included in the simulation to determine how the results of the simulation vary.

The active region to be simulated in this study is AR 11437. The evolution of AR 11437 has previously been studied by \citet{Yardley-2017}, where 20 bipolar active regions were analysed in order to investigate the role of flux cancellation in the production of CMEs. They found that a combination of shear, convergence and cancellation is required to build a pre-eruptive magnetic structure. This is consistent with the \citet{vB-1989} scenario. AR 11437 produced three eruptions during the time period studied. Two of the ejections occurred during the active region emergence phase, originating from an external PIL formed by the active region periphery and quiet Sun magnetic field. The final eruption occurred during the decay phase and was produced at the internal PIL of the active region. All three eruptions took place after flux cancellation had occurred either along the internal or external PIL. The three eruptions that were observed in the 193 \AA\ \textit{SDO}/AIA channel had no observable signatures in the white-light coronograph data, therefore it remains uncertain whether the material was ejected into interplanetary space.


This paper is structured as follows. Section \ref{sec:obs} describes the photospheric and coronal observations of AR 11437, Section \ref{sec:prop} details the properties of the AR including the evolution of magnetic flux and tilt angle. Section \ref{sec:sim} outlines the simulation method including the photospheric boundary conditions used. Section \ref{sec:res} gives the results and a comparison of the simulations when global parameters are varied and finally Section \ref{sec:con} discusses the results and concludes the study.

\section{Observations} \label{sec:obs}

\subsection{Photospheric Magnetic Field Evolution}

\begin{figure*}[t]
\epsscale{1.1}
\plotone{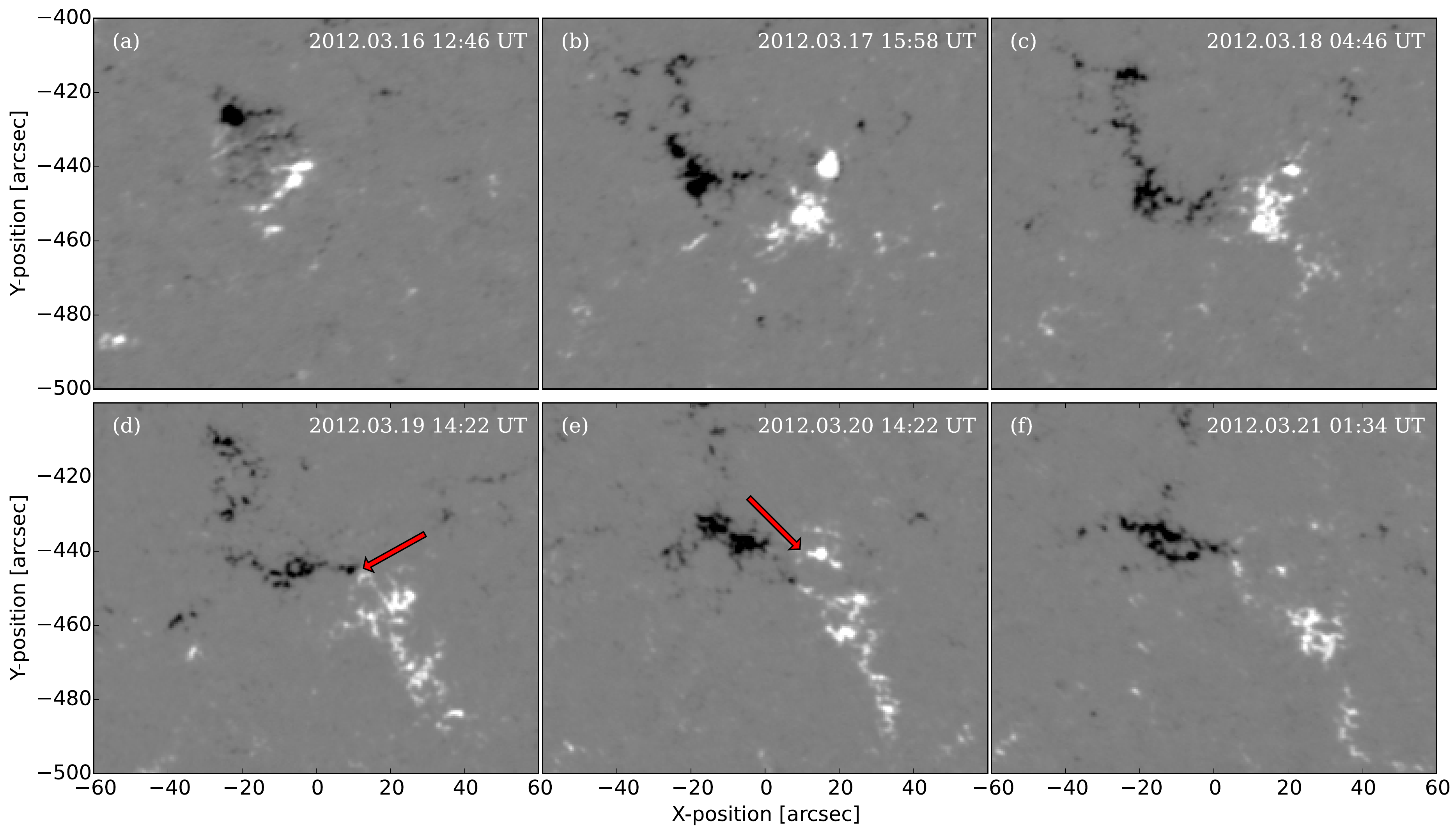}
\caption{The photospheric field evolution of AR 11437 as shown by a time sequence of \textit{SDO}/HMI LoS magnetograms from 2012 March 16--21. The white (black) contours represent the positive (negative) polarities corresponding to saturation levels of $\pm$500 G. The red arrows indicate the two sites of small-scale flux emergence. The magnetograms have been de-rotated to disk centre (2012 March 19 at 09:00 UT). \label{fig1}}
\epsscale{1}
\end{figure*}

The photospheric magnetic field evolution of AR 11437 is studied during the time period beginning 2012 March 16 12:46 UT until 2012 March 21 01:34 UT using the 720 s data series \citep{Couvidat-2016} produced by the Helioseismic Magnetic Imager (HMI: \citealt{Schou-2012}) on board the \textit{Solar Dynamics Observatory (SDO: \citealt{Pesnell-2012})}. The time period captures the full evolution of the AR allowing AR 11437 to be analysed from emergence to decay (see Figure \ref{fig1}).

AR 11437 emerges on 2012 March 16 into the Sun's southern hemisphere. The AR has a simple bipolar configuration where the polarities are initially aligned north-south (Figure \ref{fig1} (a)). The leading positive polarity is further from the equator than the following polarity, such that the bipole is anti-Joy's Law. During the first two days of observations the AR remains in it's emergence phase and rotates counter-clockwise. The rotation continues until the bipole is aligned east-west with a Hale orientation. This aspect of rotation has also been seen in flux emergence simulations \citep{Syntelis-2017}. The region reaches its peak unsigned magnetic flux on 2012 March 17 at 15:58 UT (Figure \ref{fig1} (b)), where the unsigned magnetic flux is defined as half the sum of the total positive and negative flux. The AR then enters its decay phase, starts to disperse and small-scale magnetic features converge towards the internal PIL. This leads to flux cancellation along the PIL between March 17--19 (Figure \ref{fig1} (b--d)). On 2012 March 18 the AR starts to exhibit a strong counter-clockwise rotation (Figure \ref{fig1} (d)). AR 11437 then crosses central meridian on 2012 March 19 at 09:00 UT. Small episodes of emergence are observed during the final two days of the evolution on 2012 March 19 at 10:00 UT and March 20 at 14:00 UT (red arrows in Figure \ref{fig1} (d--e)). The AR continues to disperse until the leading positive polarity is more diffuse than the following negative polarity (Figure \ref{fig1} (f)).

\subsection{Coronal Evolution}

\begin{figure*}[t]
\epsscale{0.8}
\plotone{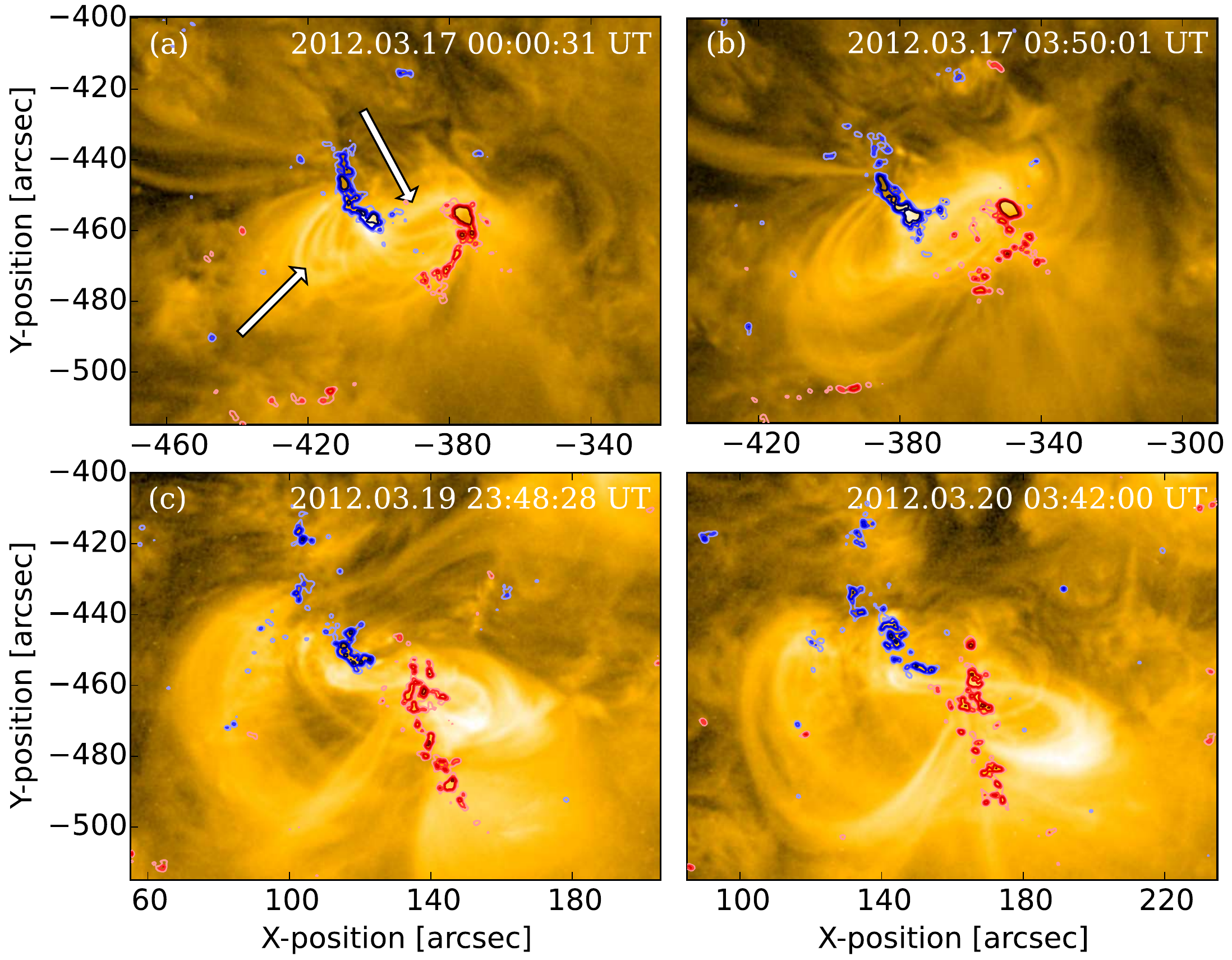}
\caption{The coronal evolution of AR 11437 observed using high-resolution 171 \AA\ images taken from \textit{SDO}/AIA. The LoS magnetic field from \textit{SDO}/HMI is shown at a saturation of $\pm$100 G with the red (blue) contours corresponding to positive (negative) magnetic flux, respectively. The white arrows in panel (a) indicate the presence of non-potential coronal loops, which during the time period analysed evolve to become more sheared as seen in panel (b). In the later stages of the coronal evolution J-shaped loops are observed to develop, shown in panels (c) and (d). \label{fig2}}
\epsscale{1}
\end{figure*}

\begin{figure}

\plotone{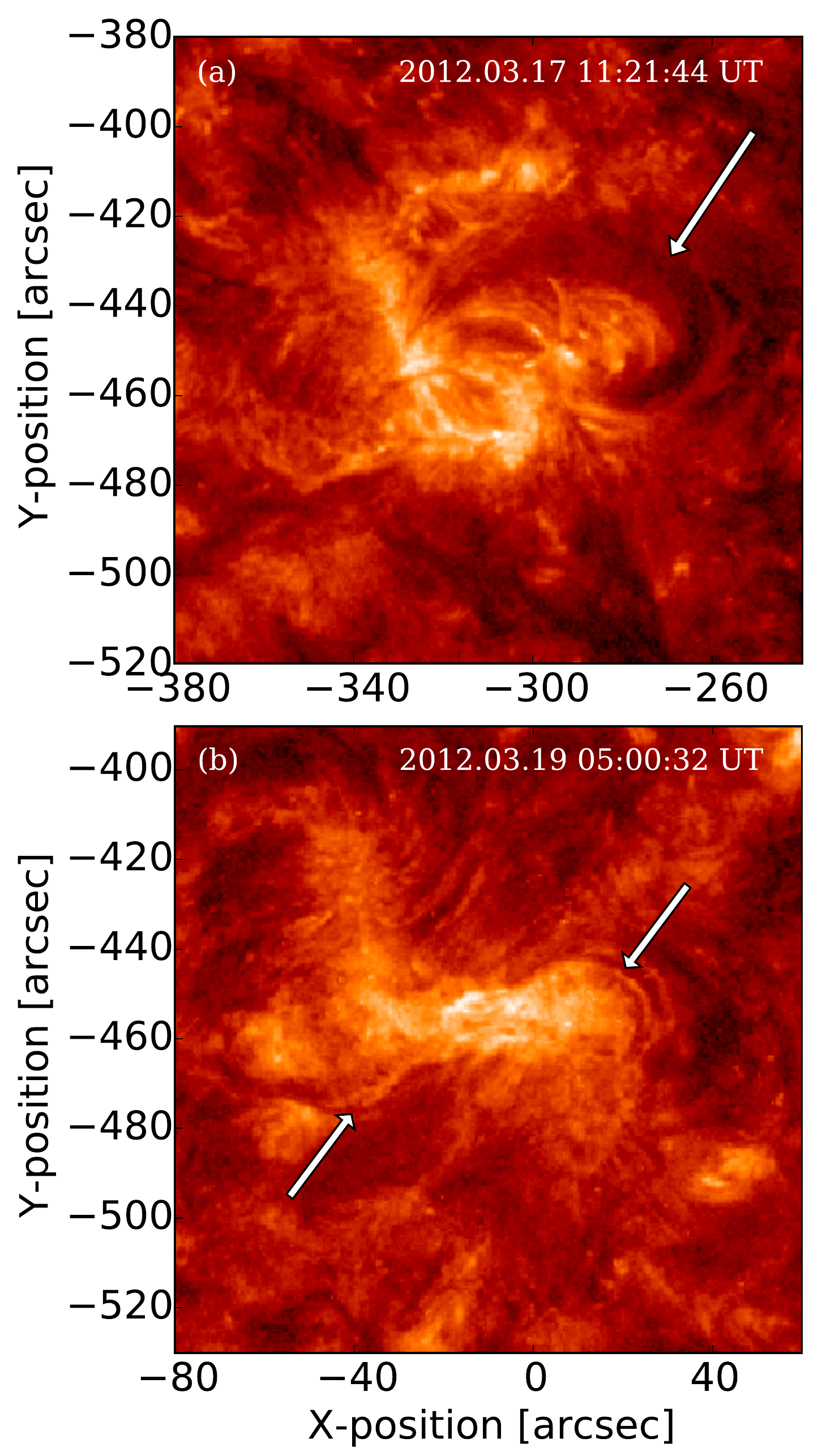}
\epsscale{0.75}
\caption{Filament formation in AR 11437 observed using high-resolution 304 \AA\ images from \textit{SDO}/AIA. The filament that forms in the north-west of the AR, shown in panel (a), is observed to erupt but reforms after the eruption. Panel (b) shows that a second filament forms in the south-east of the active region, which extends out into the quiet Sun. The white arrows in both panels show the locations of the two filaments. \label{fig3}}
\epsscale{1}
\end{figure}

The coronal evolution of AR 11437 is analysed using EUV images taken by the Atmospheric Imaging Assembly (AIA: \citealt{Lemen-2012}) on board \textit{SDO}, which provides full-disk, high-resolution observations in three UV continuum wavelengths and seven EUV bandpasses. These observations have a spatial and temporal resolution of 1.5" and 12 s, respectively. We focus on the 304 and 171 \AA\ passbands, which are dominated by plasma emission at temperatures of approximately 0.05 and 0.6 MK.

During the early stages of evolution the coronal loops, observed using the 171 \AA\ passband, evolve quickly from a potential (not shown) to a highly sheared configuration. The sheared coronal loops are highlighted by white arrows in Figure \ref{fig2} (a). These sheared loops periodically brighten as emergence continues and a system of dark loops form at the periphery of the positive polarity sunspot (Figure \ref{fig2} (b)). On 2012 March 17 there are two eruptions that take place in quick succession during the emergence phase. There are no soft X-ray flares associated with these eruptions although, small brightenings accompany both ejections. While there are signatures present in the low corona that suggest these ejections are CMEs, there is no clear evidence of any of these eruptions in the white-light coronograph data. This suggests that the CMEs may have a low density or that they may be confined or failed eruptions. Therefore, we will refer to these events as either eruptions or ejections rather than CMEs. 

The first eruption occurs on 2012 March 17 at approximately 05:09 UT when the dark loop system that is located to the west of the positive polarity sunspot erupts (Figure \ref{fig2} (b)). A filament is observed to form in 304~\AA\ (white arrow in Figure \ref{fig3} (a)) which erupts a few hours later on 2017 March 17 at approximately 10:45 UT.
On 2012 March 18 the coronal loops slowly start to reform. Two filaments form in 304 \AA\ located in the north-west and south-east of the AR (white arrows in Figure \ref{fig3} (b)). In the coronal emission a J-shaped structure becomes visible on 2012 March 19 (Figure \ref{fig2} (c)). This structure, which is best observed in 171 \AA, is not always seen. The J-shaped loops disappear and different loops that are part of the same structure become visible (Figure \ref{fig2}(d)). The third and final ejection from this region occurs on 2012 March 20 at 14:31 UT when the J-shaped loop system erupts and post-reconnection loops are observed.

\section{AR Properties} \label{sec:prop}

The coronal field evolution of AR 11437 is simulated using a continuous time sequence of lower boundary conditions that are generated from photospheric LoS magnetograms (as discussed in Section \ref{sec:boundary}). Before the simulations are carried out a clean-up process is applied to the magnetograms, which is described in Appendix \ref{sec:A}. The cleaning procedure includes time-averaging, removal of low flux values and small magnetic features along with flux balancing. This is to remove quiet sun magnetic features whilst ensuring that the large-scale evolution of the AR is retained. This section describes the properties of AR 11437 that are derived from the cleaned magnetograms.

\subsection{Magnetic Flux Evolution}

\begin{figure}
\epsscale{1.15}
\plotone{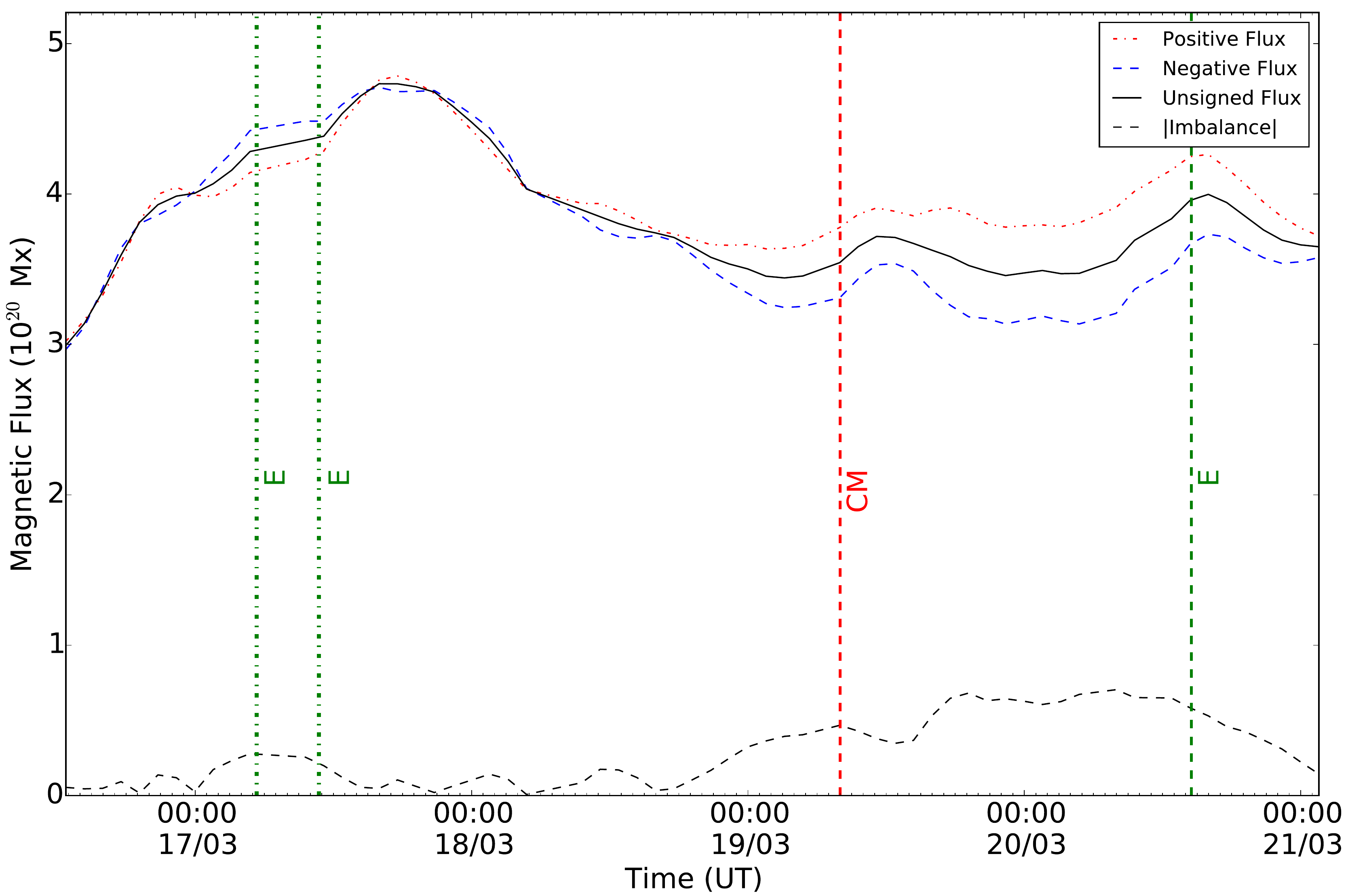}
\caption{Positive (red dot-dashed line), negative (blue dashed line) and unsigned (black solid line) magnetic flux of AR 11437 between 2012 March 16 12:46 UT and 2012 March 21 01:34 UT. The black dashed line is the absolute value of the flux imbalance of the AR as a function of time. The green dot-dashed lines show the times of the eruptions (E, 2012 March 17 05:09 UT, 2012 March 17 10:45 UT and 2012 March 20 at 14:21 UT) and the red dashed line represents the central meridian passage time (2012 March 19 at 08:00 UT). \label{fig4}}
\epsscale{1}
\end{figure}

The magnetic flux variation and the absolute flux imbalance for AR 11437 is shown in Figure \ref{fig4} for the time period beginning 2012 March 16 12:46 UT until 2012 March 21 01:34 UT. The evolution of the magnetic flux after the cleaning process is applied remains the same as in the raw data. There is a flux imbalance present during both the emergence and decay phase of the AR. During these two phases first negative and then positive magnetic flux dominates. This is a geometric effect caused by the presence of an east-west horizontal component in the magnetic field that links the two AR polarities \citep{Green-2003}. The strong horizontal component has an additional contribution to the LoS magnetic flux with the imbalance increasing with distance from central meridian. The flux increases in the polarity closest to the solar limb and so this effect reverses when the AR crosses central meridian. While this effect exists the flux imbalance is small compared to the total flux throughout the time period considered.

AR 11437 emerges onto the disk on 2012 March 16 into a region of relatively evenly distributed quiet Sun positive and negative polarity magnetic field in the southern hemisphere. The magnetic flux continues to increase for the first two days of observations. During the flux emergence phase two eruptions occur on 2012 March 17 at 05:09 UT and 10:45 UT, which are represented by the green dashed lines in Figure~\ref{fig4}. The region reaches a peak unsigned magnetic flux of 4.7 $\times$ 10$^{20}$ Mx on 2012 March 17 at 15:58 UT. The magnetic flux is observed to decrease between 2012 March 17 at 15:59 UT until 2012 March 19 03:11 UT as flux cancellation occurs along the internal PIL. Approximately, 1.3 $\times$ 10$^{20}$ Mx of magnetic flux is cancelled, calculated using the cleaned magnetograms, which amounts to 27 \% of the total unsigned AR magnetic flux. This can be compared to the results of a study by \citet{Yardley-2017}, which found a total of 1.7 $\times$ 10$^{20}$ Mx is cancelled during this time period, amounting to 31 \% of the total unsigned active region flux. The difference between these two values is due to the different clean-up processes that have been applied to the magnetograms along with the area that is used to calculate the magnetic flux in each study. In the present study the magnetograms are treated with a number of processes including time-averaging, low flux removal and isolated feature removal. All processes are carried out so that the magnetograms can be used as lower boundary conditions in the coronal field simulations. In contrast, only smoothing is applied in the observational study so that magnetic features that are not part of the active region are disregarded. Therefore, the only contribution to the calculation of magnetic flux is from the active region magnetic features. However, in the simulation the magnetic flux is calculated from the reduction in magnetic flux using the entire field of view of the magnetograms. 

On 2012 March 19 at 10:00 UT and 2012 March 20 at 14:00 UT there are two small episodes of emergence at the internal PIL. Shortly after the final episode of flux emergence there is a third eruption at around 14:31 UT on 2012 March 20.

\subsection{Tilt Angle}

\begin{figure}
\epsscale{1.15}
\plotone{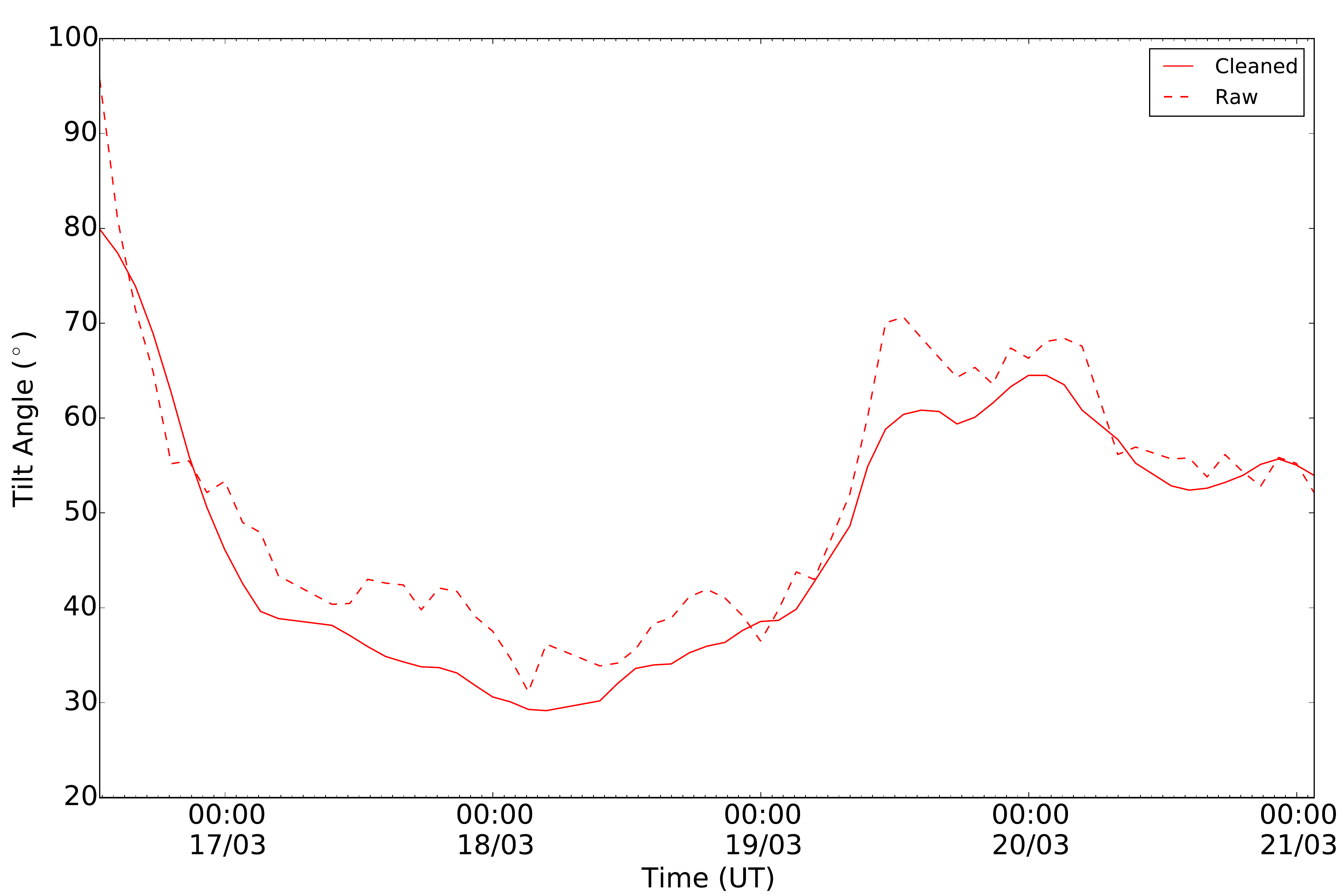}
\caption{The evolution of the tilt angle as a function of time for AR 11437. The red dashed (solid) line represents the raw (clean) tilt angle of the AR. \label{fig5}}
\epsscale{1}
\end{figure}

The variation in tilt angle with time is calculated during the same time period as the evolution of the magnetic flux (2012 March 16 12:46 UT -- 2012 March 21 01:34 UT). The tilt angle of an active region is defined as the angle between the east-west direction on the Sun and the line that connects the centre of flux of the positive and negative AR polarities. Appendix \ref{sec:B} describes how the tilt angle of the AR is calculated. Figure \ref{fig5} shows that the evolution of the tilt angle before (red dashed line) and after (red solid line) the clean-up process is applied does not change. 

At the start of the emergence phase the AR polarities have a tilt angle of roughly 80$^{\circ}$ as the line that connects the flux-weighted centres of the active region polarities is almost aligned north-south. As AR 11437 continues to emerge the polarities rotate counter-clockwise until the active region is aligned east-west with a tilt angle of approximately 30$^{\circ}$ on 2012 March 18. During the decay phase, the active region polarities rotate clockwise, with the tilt angle increasing to around 65$^{\circ}$ on 2012 March 20. On the final day of observations, when the magnetic field is more dispersed, the active region begins to rotate counter-clockwise again and the tilt angle decreases slightly. The variation of the tilt angle is due to the forces that are generated during the emergence and decay of the active region and are opposite in nature to that expected from differential rotation.

\section{The Simulation} \label{sec:sim}

\subsection{Coronal Field Evolution}

The coronal magnetic field evolution of AR 11437 is simulated by using a magnetofrictional relaxation technique to generate a continuous time series of NLFF magnetic fields from a time sequence of photospheric LoS magnetograms. The method of magnetofrictional relaxation was originally proposed by \citet{Yang-1986} and has since been successfully applied in numerous studies of filaments \citep{vB-2000, Mackay-2003, Mackay-2005, Mackay-2009}, and magnetic flux rope formation \citep{Mackay-2006a, Gibb-2014}. As these simulations evolve the coronal magnetic field through a continuous series of NLFF equilibria, magnetic connectivity and flux are preserved. This allows the injection of magnetic energy and helicity into the corona to be analysed. When using independent coronal field extrapolations this type of analysis is not possible as with independent extrapolations these quantities are not preserved from one time to the next. 

To analyse the coronal field evolution of AR 11437, we use the same coronal modelling technique as \citet{Mackay-2011}.
The evolution of the 3D magnetic field $\mathbf{B}$, where $\mathbf{B} = \nabla \times \mathbf{A}$, is governed by the induction equation,

\begin{equation}
\frac{\partial \mathbf{A}}{\partial t}  = \mathbf{v} \times \mathbf{B} ,
\end{equation}
where $\mathbf{A}$ is the magnetic vector potential and $\mathbf{v}$ is the magnetofrictional velocity. In the model, an artificial frictional term known as the frictional coefficient $\nu^{'}$, is included in the equation of motion, which under the force-free approximation (steady state, neglecting any external forces) reduces to

\begin{equation}
\mathbf{j} \times \mathbf{B} - \nu' \mathbf{v} = 0,
\end{equation}
where $\mathbf{j} = \nabla \times \mathbf{B}$. Hence, the magnetofrictional velocity $\mathbf{v}$, can be expressed as

\begin{equation}
\mathbf{v} = \frac{1}{\nu'}  \mathbf{j} \times \mathbf{B},
\end{equation}
where the frictional coefficient takes the form $\nu' = \nu B^{2}$. The magnetofrictional velocity acts to ensure that the magnetic field remains close to a force-free equilibrium as the field is perturbed via boundary motions. The frictional coefficient is set such that, $\nu = 3000$~km$^{2}$~s$^{-1}$, as this value has produced the best match between the simulated coronal field and the coronal observations in previous studies. A staggered grid is used in the computations to calculate the variables $\mathbf{A}, \mathbf{B}$ and $ \mathbf{j}$ to second-order accuracy. The computational domain represents the solar corona and the bottom of the box represents the photosphere. The lower boundary conditions are provided by the observed LoS magnetograms, which undergo various clean-up processes (see Appendix \ref{sec:A}), namely time-averaging, removal of isolated features and low flux values. Closed boundary conditions are used for the sides of the box whereas, the top of the domain may have either open or closed boundaries. If open boundary conditions are selected then the magnetograms need not be flux balanced. In contrast, if closed boundary conditions are used then the magnetograms have to be flux balanced to ensure that $\nabla \cdot \mathbf{B} = 0$ is satisfied in the coronal volume. In this study, simulations are carried out using both open and closed top boundaries to determine the relative effect on the evolution of the coronal field (see Section \ref{sec:param}). The generation of the lower boundary conditions and initial condition are discussed in the following section.

\subsection{Photospheric Boundary Conditions and Initial Condition} \label{sec:boundary}

To simulate the evolution of AR11437, 62 LoS magnetograms from the HMI 720 s data series are used with a cadence of 96 minutes. The medium cadence represents a middle ground between the cadence available from current space-borne and ground-based missions and future instruments, which may have limited telemetry or temporal resolution. The magnetograms span a 4.5 day period around the central meridian passage of the AR and are cleaned using the clean-up process described in Appendix \ref{sec:A}. 

A number of simulations have been carried out with a variety of parameters and open or closed top boundary conditions as described in Section \ref{sec:param}. The simulations have a lower resolution than the LoS magnetograms as the 278 $\times$ 279 pixel magnetograms are interpolated onto a grid size of 256$^{2}$. A continuous time sequence of lower boundary conditions are generated from the corrected magnetograms, which are designed to match the cleaned LoS magnetograms, pixel by pixel, every 96 minutes.

To model the coronal field evolution of AR 11437, the horizontal components ($A_{xb}, A_{yb}$) of the magnetic vector potential $\mathbf{A}$ on the base corresponding to each magnetogram must be determined. This process is carried out as follows

\begin{enumerate}
\item Each of the observed LoS magnetograms, $B_{z}(x, y ,k)$ for $k = 1 \rightarrow 62$ are taken, where $k$ represents the discrete 96 minute time index.

\item The horizontal components of the vector potential $\mathbf{A}$ at the base ($z=0$) are expressed in the form
\begin{eqnarray}
A_{xb} (x, y, k) & = & \frac{\partial \Phi}{ \partial y}, \\
A_{yb} (x, y, k) & = &  - \frac{\partial \Phi}{\partial x}, 
\end{eqnarray}
where $\Phi$ is a scalar potential.

\item For each discrete time index $k$, the following equation
\begin{equation}
B_{z} = \frac{\partial A_{yb}}{\partial x} - \frac{\partial A_{xb}}{\partial y},
\end{equation}

then becomes
\begin{equation}
\frac{\partial^{2} \Phi}{\partial x^{2}} + \frac{\partial^{2} \Phi}{\partial y^{2}} = - B_{z}, \label{eq:7}
\end{equation}
\end{enumerate}
which is solved using a multigrid numerical technique (\citealt{Finn-1994, Longbottom-1998} and references therein). In the construction of $A_{x}$ and $A_{y}$ from $B_{z}$ in the 2D plane identical boundary conditions are chosen from one time to the next based on the Coulomb gauge. This ensures that the change in $A_{x}$ and $A_{y}$ used to drive the simulation is minimised within the Coulomb gauge. Full details and be found in the papers of \cite{Mackay-2009, Mackay-2011}, based upon the paper of \cite{Finn-1994}. In the recent paper of \cite{Yeates-2017} a new localised technique for determining the boundary condition at the level of the photosphere has been put forward. Future studies will consider the consequences of this.

By solving Equation \ref{eq:7} for the scalar potential $\Phi$, the horizontal components of the vector potential on the base ($A_{xb}, A_{yb}$) can be determined for each discrete time interval, 96 minutes apart. For a continuous time sequence between each of the observed distributions to be produced, the horizontal components $A_{xb}$ and $A_{yb}$ are linearly interpolated between each time interval $k$ and $k+1$. The fields are interpolated using 500 interpolation steps. This effectively evolves the magnetic field from one observed photospheric field to the next. Through using this process, additional numerical techniques, such as local correlation tracking, are not required as the horizontal velocity does not need to be determined. Furthermore, the removal of undesirable effects such as numerical overshoot or magnetic flux pile up at cancellation sites is not necessary as these numerical effects do not occur. 

Due to the numerical technique described above, there are two timescales involved in the evolution of the lower boundary condition. The first is the 96 minute timescale between observations, the second is a timescale of 11.52 s, introduced to produce the advection of the magnetic polarities between observed states by interpolation, along with the relaxation of the coronal field. The process described above reproduces the cleaned LoS magnetograms with a discrete time interval of 96 minutes, providing a description of the magnetogram observations that is highly accurate.

\section{Results} \label{sec:res}

\subsection{Magnetic Field Line Evolution} \label{sec:closed_results}

The evolution of the simulated coronal magnetic field will now be discussed for the simplest case where closed boundary conditions are used at the top of the computational box. The results found for both open and closed top boundary conditions were very similar and will be discussed in Section~\ref{sec:param}. Therefore the nature of the upper boundary condition does not have a significant effect on the evolution of the field.

\begin{figure}
\epsscale{1.15}
\plotone{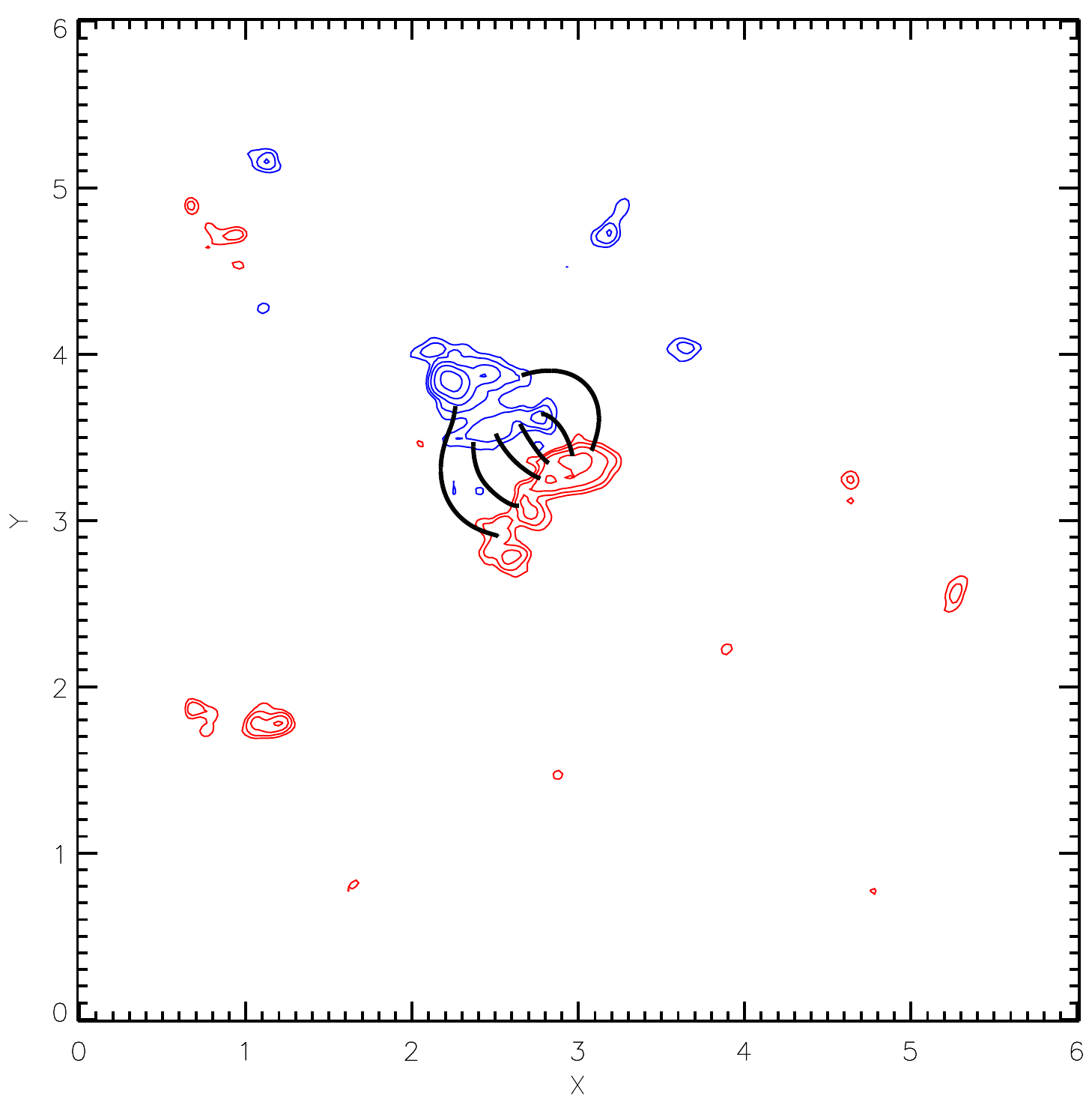}
\caption{A selection of magnetic field lines (black) that illustrate the initial potential field condition. The red (blue) contours represent the positive (negative) photospheric magnetic field. \label{fig6}}
\end{figure}

The initial potential field condition for the coronal field is constructed from the first LoS magnetogram taken on 2012 March 16 at 12:46 UT and is shown in Figure \ref{fig6}. The AR has a simple, bipolar configuration where the field lines that connect between the positive (red contours) and negative (blue contours) polarities consist of semi-circular loops. The simulated field lines match well with the coronal loops observed at this time.

\begin{figure*}[t]
\epsscale{0.7}
\plotone{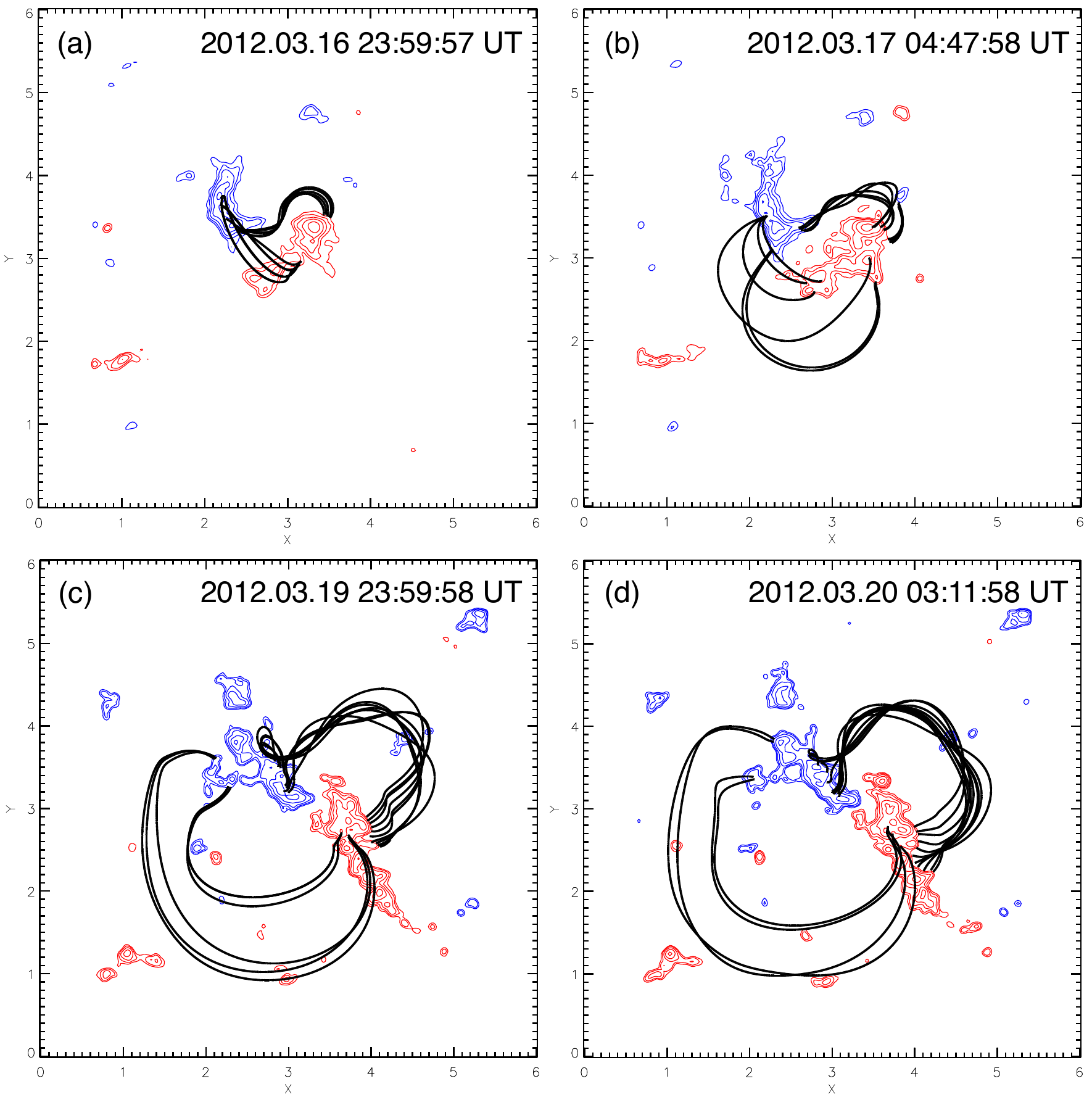}
\caption{A series of field line plots taken from the simulation that roughly correspond to the timings of the AIA observations shown in Figure \ref{fig2}. The positive and negative polarities are represented by the red and blue contours, respectively. \label{fig7}}
\end{figure*}

Figure \ref{fig7} shows a series of field line plots taken from the simulation that approximately correspond to the times of the 171 \AA\ AIA observations in Figure \ref{fig2}. A selection of field lines are displayed to demonstrate that within the simulation many of the observable features are also present. For example, in panel (a) of Figure \ref{fig7}, the simulated magnetic field lines that are located along the internal PIL in the north of the active region are S-shaped, indicating the presence of shear. A similar sheared feature is also present in the coronal observations (see Figure \ref{fig2} (a)). While there is a good agreement between the observations and simulations there are however some differences. One key difference is that in the coronal emission observations the S-shaped feature appears as one continuous structure that extends to the south of the active region with its eastern footpoint located at the periphery of the negative polarity. This is not the case for the simulation as the S-shaped structure only partially extends along the internal PIL and the endpoints are fixed in the center of the negative polarity sunspot. However, care must be taken in the direct comparison between the observed coronal loops and field lines as the loops may represent integrated structures along the LoS rather than single field lines. The simulated field lines that are located to the south of the S-shaped structure have a potential configuration, most likely due to using a potential field initial condition. 

In panel (b) of Figure \ref{fig7} the S-shaped field lines have evolved into a flux rope configuration at the same location as the sheared coronal loops visible in Figure \ref{fig2} (b). As found in the previous panel the simulated flux rope does not extend as far south as the sheared coronal loops. The arcade field lines that are situated directly west of the simulated flux rope are in good agreement with the dark loop system present in the observations.

The rapid evolution of the magnetic field in the simulation at this time suggests that the flux rope starts to rise and erupt between the current and previous time steps. Signatures of the eruption in the simulation include the deformation of the flux rope and post-reconnection field lines forming below the rope. The magnetofrictional simulation does not capture the full dynamics of the eruption as the flux rope is confined in the low corona. The flux rope does not rise to the top of the box even when open top boundary conditions are used, suggesting that the eruption may be confined. The eruption takes place approximately 1 hour before the ejection of the dark loop system that is seen in the 171 \AA\ observations. It is important to note that there are no white-light signatures associated with the eruption. While in the simulation the eruption occurs roughly 1 hour before that found in the observations, we note that to drive the evolution of the coronal field, magnetogram data with a 96 minute cadence is used. Therefore, the eruption occurs within the time resolution of the boundary data. A more detailed comparison between the timings of the flux rope eruptions in the simulation and the observations of ejections will be discussed in the Section \ref{sec:mfe}. Also, the field lines that are located in the south of the active region, below the flux rope, are now relatively similar to the loops seen in the coronal emission. Figure \ref{fig7} panel (c) shows the flux rope three days after the first two eruptions are observed. The field lines of the simulated flux rope have grown in length resulting in a larger structure. This is consistent with the J-shaped loop structure observed in Figure \ref{fig2} (c). The footpoints of the simulated flux rope that are rooted in the center of the negative polarity are shifted to the west and are closer to the internal PIL compared to the footpoints of the J-shaped loops in the 171 \AA\ observations. The simulated loops in the south of the active region are in good agreement with the observations. In Figure \ref{fig7} (d) some of the footpoints of the simulated flux rope are situated further south, along the periphery of the positive polarity magnetic field. This is consistent with the evolution of the J-shaped emission in the observations (Figure \ref{fig2} (d)). At this time the simulated flux rope is in the process of erupting again. The eruption in the simulation occurs roughly 10 hours before the third eruption is observed (see Section \ref{sec:param}).

The simulated field lines in Figure \ref{fig7} are in good agreement with the 171 \AA\ observations. This suggests that by using the magnetofrictional relaxation technique of \citet{Mackay-2011} to produce a continuous time series of NLFF magnetic fields, driven by photospheric LoS magnetograms, it is possible capture observable features of AR 11437. This is particularly true in the north of the active region where the formation, evolution and eruption of the simulated flux rope agrees well with the coronal observations. However, the field lines located in the south of the active region deviate from the observed coronal emission during the early phases of active region evolution. This deviation may be due to the choice of the initial potential condition and closed top boundary conditions. The discrepancy between the simulation and observations improves during the later stages of active region evolution. The effect of additional global parameters, the initial condition and an open top boundary on the simulated coronal field are explored in Section \ref{sec:param}.

\subsection{Parameter Study} \label{sec:param}

In the simulation there are three additional terms that can be implemented by modifying the induction equation as follows

\begin{equation}
\frac{\partial \mathbf{A}}{\partial t} = \mathbf{v} \times \mathbf{B} - \eta \mathbf{j} + \frac{\mathbf{B}}{B^{2}} \nabla \cdot (\eta_{4} B^{2} \nabla \alpha) + \mathbf{H},
\end{equation}

where $\alpha$ is given by

\begin{equation}
\alpha = \frac{\mathbf{B} \cdot \nabla \times \mathbf{B}}{B^{2}}.
\end{equation}

The second term on the right-hand-side represents Ohmic diffusion, which simply originates from the resistive form of the induction equation, where $\eta$ is the resistive coefficient. 

The third term is known as hyperdiffusion \citep{Boozer-1986, Strauss-1988, Bhattacharjee-1995} and includes the coefficient of hyperdiffusion $\eta_{4}$. Hyperdiffusion is an artificial diffusion that is used to smooth out gradients present in the force-free parameter $\alpha$, whilst allowing the conservation of total magnetic helicity \citep{vB-2007}. The final term represents an additional injection of helicity at the photosphere. In particular, the injection of magnetic helicity could be due to torsional Alfv{\'e}n waves propagating into the corona from below the surface or as a result of small-scale vortical motions that are associated with granular or supergranular cells \citep{Antiochos-2013}. This helicity injection $\mathbf{H}$, can be expressed through the source term

\begin{equation}
\mathbf{H} = - \nabla_{z} (\zeta B_{z}),
\end{equation}

where $\zeta$ is the helicity injection parameter and parameterises the rate and scale of helicity injection at the photospheric surface. Helicity injection can introduce strongly sheared field when horizontal motions in the photosphere do not inject enough helicity into the coronal field. For a more detailed description and derivation of this term refer to \citet{Mackay-2014}.

\begin{deluxetable}{ccc} 
\tabletypesize{\scriptsize} 

\tablewidth{0pt}
\tablecolumns{3} 
\tablecaption{Parameter Study \label{tab:table1}} 
\tablehead{ \colhead{Simulation} & \colhead{Boundary \&} & \colhead{Additional} \\ \colhead{No.} & \colhead{Initial Conditions} & \colhead{ Terms}}
\startdata 
1 & Closed & - \\
2 & Open & - \\
3 & Closed, $\alpha = 4.9 \times 10^{-8}$ & - \\
4 & Closed & $\eta=25$ \\
5 & Closed & $\eta=50$ \\
6 & Closed & $\eta=100$ \\
7 & Closed & $\eta_{4}=100$\\
8 & Closed & $\eta_{4}=200$ \\
9 & Closed & $\zeta = -1$ \\
10 & Closed & $\zeta = -10$ \\
\enddata 

\tablecomments{The boundary conditions, initial conditions and additional terms that are used to conduct the simulations. The top boundary conditions of the simulation can either be open or closed. Depending on the boundary conditions the simulations can have either a potential or linear force-free initial condition where the force-free parameter $\alpha$ is given in units of m$^{-1}$. Ohmic diffusion and helicity injection are included using $\eta$ and $\zeta$, which both have units of km$^{2}$s$^{-1}$, whereas, the coefficient of hyperdiffusion has units of km$^{4}$s$^{-1}$.}
\end{deluxetable}

\subsubsection{Magnetic Field Evolution} \label{sec:mfe}

We now conduct a parameter study where the top boundary, initial condition and non-ideal terms are varied in the simulation. This is to analyse the effect that these  conditions have on the coronal field evolution described in Section \ref{sec:closed_results}. In total, ten simulations are performed (see Table \ref{tab:table1}) using either closed or open boundary conditions. If the top boundary is closed a LFF initial condition can be used where the force-free parameter $\alpha$ can take a range of values between [-1.6, 1.6] $\times$ 10$^{-7}$m$^{-1}$. Additional global parameters are also included such as Ohmic diffusion, hyperdiffusion and helicity injection. We find that the addition of these parameters does not significantly affect the overall coronal evolution of the simulated magnetic field. However, when Ohmic diffusion is included the simulated field has less twist than in the other cases. When the value of this term is increased the magnetic twist decreases further. This is expected as by adding coronal diffusion into the simulation this decreases the twist by decreasing the amount of poloidal flux. 

\begin{figure}
\plotone{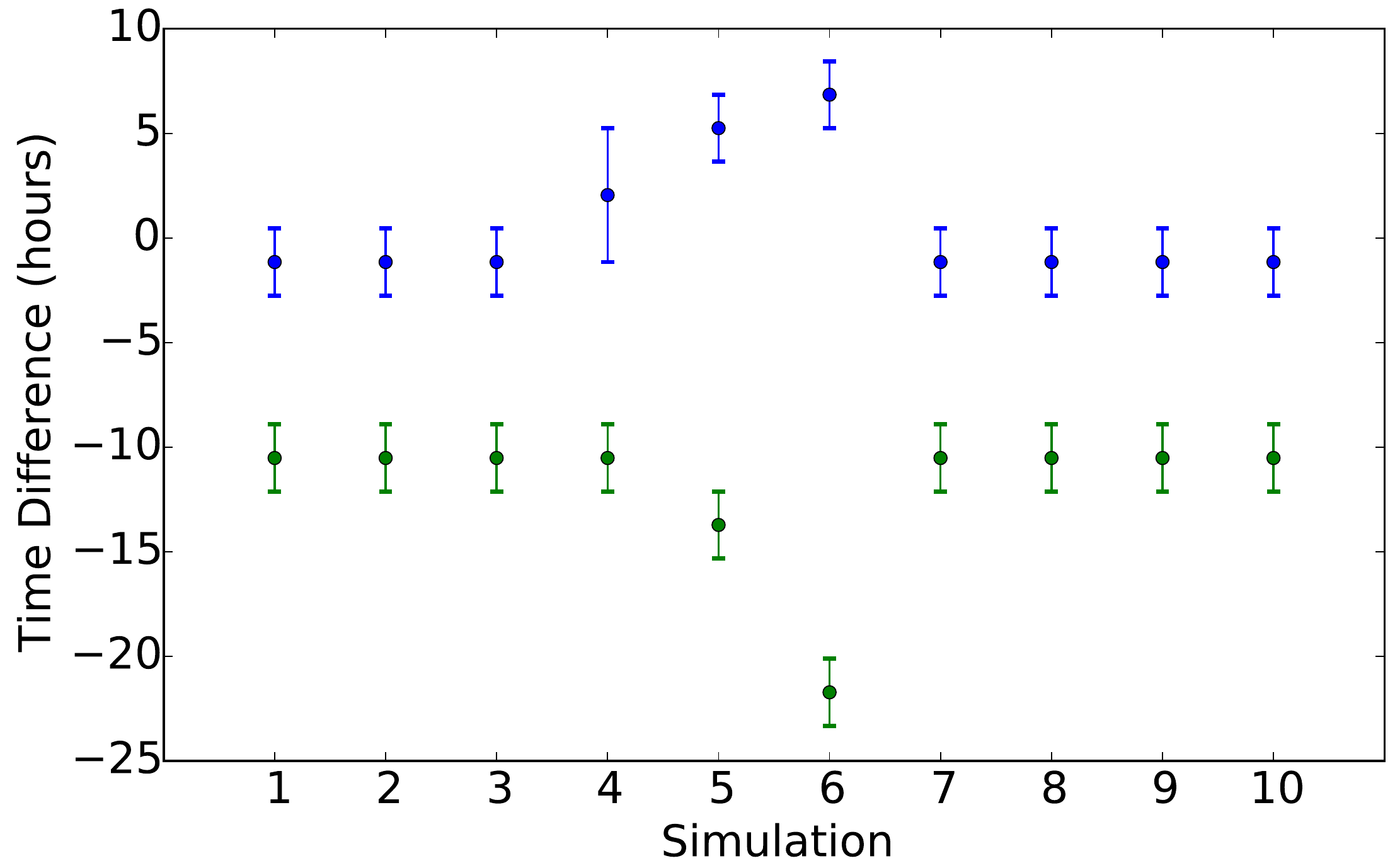}
\caption{The time difference in hours between the two flux rope eruptions that occur in the ten simulations and the eruptions seen in the 171 \AA\ observations on 2012 March 17 at 05:09 UT (blue) and 2012 March  20 at 14:31 UT (green). As the flux rope eruptions occur between two time steps in the simulation the points represent the time difference taken between the observed eruptions and the central time between the two time steps. The error bars represent the error in time difference due to the resolution of the boundary data.  \label{fig8}}
\end{figure}

The comparison between the timings of the simulated flux rope eruptions to the ejections that occur in the observations is now discussed. As previously stated, there are three eruptions that are observed to take place on 2012 March 17 at 05:09 UT, 2012 March 17 at 10:45 UT and 2012 March 20 at 14:31 UT. The first and second eruptions are less than 6 hours apart making it impossible for the simulation to disentangle them. Therefore, we focus on the timings of the first and third ejections only. Figure \ref{fig8} shows the time difference for each simulation between the eruptions of the simulated flux ropes and the observed eruptions. The time difference between the eruptions occurring in the simulations and the observations is calculated by using the central time between the time step where the eruption has already occurred and the previous time step where there is no sign of eruption. Signatures of eruption in the simulation include kinked shaped field lines and post-reconnection loops, which form underneath the field lines of the flux rope as it rapidly rises. This time is then compared to the time of the eruption in the observations. 

For the first observed eruption, the flux rope in seven of the simulations erupts on 2012 March 17 at 04:00 UT. This is approximately 1 hour before the eruption in the observations and is within the time resolution of the photospheric boundary data. We note that when open boundary conditions are applied at the top boundary (Simulation 2) the simulated flux rope erupts in the same manner as in the closed case as it does not leave the computational box. This again suggests that this is a confined eruption. In the three remaining simulations, all of which use Ohmic diffusion, the flux rope erupts at 07:12 UT, 10:24 UT and 12:00 UT, which is roughly 2, 5 and 7 hours after the eruption in the observations, respectively. The latter times are due to Ohmic diffusion decreasing the rate of build up of twist in the flux rope. For the third and final eruption the flux rope in eight of the simulations erupts on 2012 March 20 at 04:00 UT, which is roughly 10 hours before the eruption in the observations. In the final two simulations the flux rope erupts at 00:48 UT on 2012 March 20 and 16:48 UT on 2012 March 19. This is approximately 13 and 21 hours before the observed eruption. In these simulations, where the time of the flux rope eruption deviates significantly from the other simulations, Ohmic diffusion is large taking values of 50 and 100 km$^{2}$ s$^{-1}$, respectively.

\subsubsection{Magnetic Energy}

\begin{figure}
\plotone{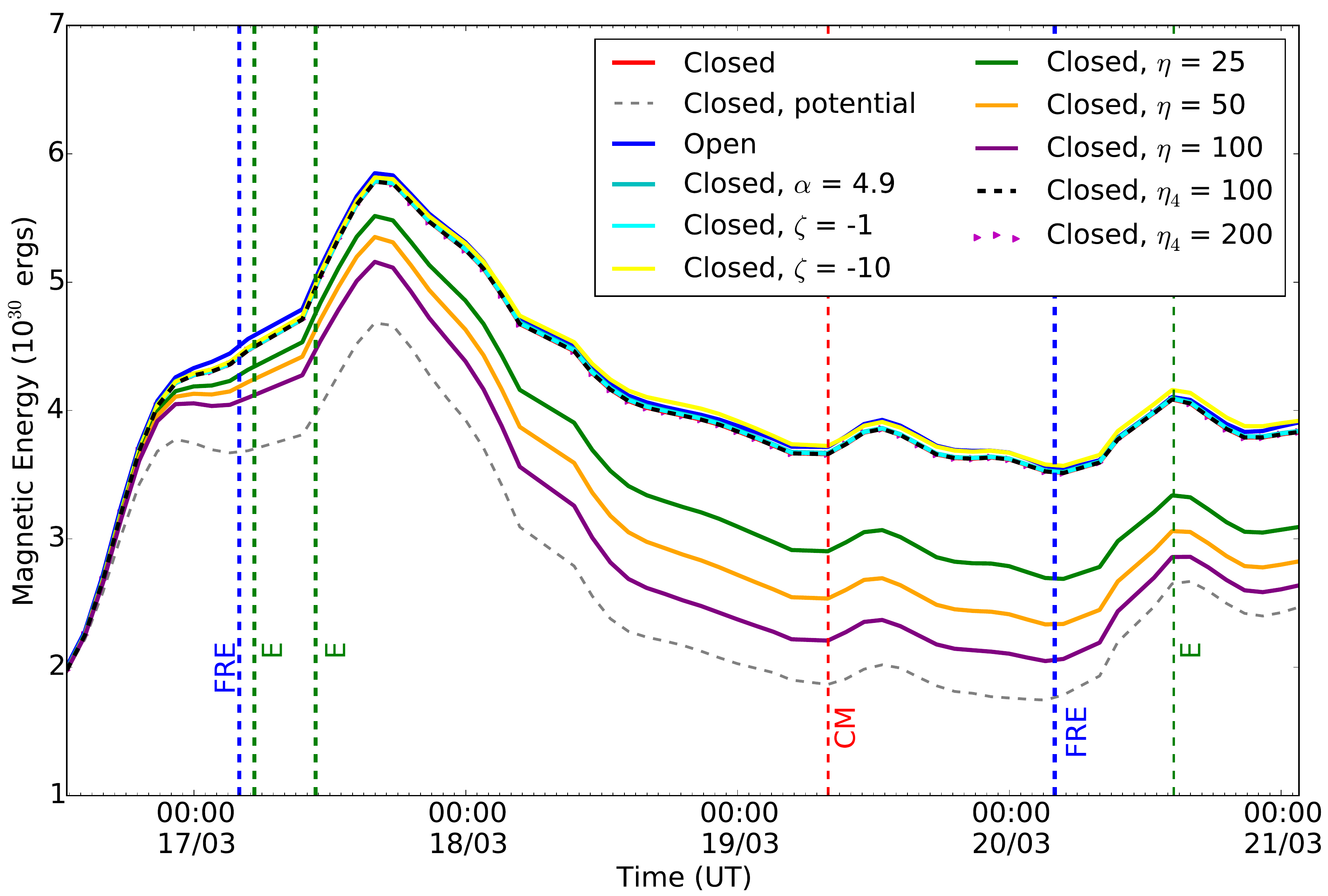}
\caption{The evolution of total magnetic energy as a function of time between 2012 March 16 12:46 UT and 2012 March 21 01:34 UT for the simulations in Table \ref{tab:table1}. The red dashed line indicates when AR 11437 crosses central meridian (CM). The blue dashed and green dot dashed lines represent the times of the simulated flux rope eruptions (FRE) and the eruptions in the observations (E), respectively. \label{fig9}}
\end{figure}

The total magnetic energy that is stored in the coronal field with time for each simulation is shown in Figure \ref{fig9}. The grey dotted line shows the total magnetic energy evolution of the potential field calculated from the same photospheric boundary conditions that are used in simulation 1 (see Table \ref{tab:table1}), which has a closed top boundary. The simulations and the potential field all start with $\sim 2 \times 10^{30}$ ergs of total magnetic energy in the system and have the same evolution throughout the time period studied. The evolution of the total magnetic energy follows that of the magnetic flux directly i.e. when there is an increase or decrease in magnetic flux due to emergence or cancellation the magnetic energy increases/decreases accordingly. While the energy evolution follows a similar trend for each simulation the total magnetic energy is systematically higher than the potential field. This is due to small-scale convective motions that evolve the magnetic elements at the photosphere and inject energy into the corona, causing the coronal field to become non-potential. For the simulations where Ohmic diffusion is incorporated the values of total magnetic energy are consistently lower. 

\subsubsection{Free Magnetic Energy}

\begin{figure}
\plotone{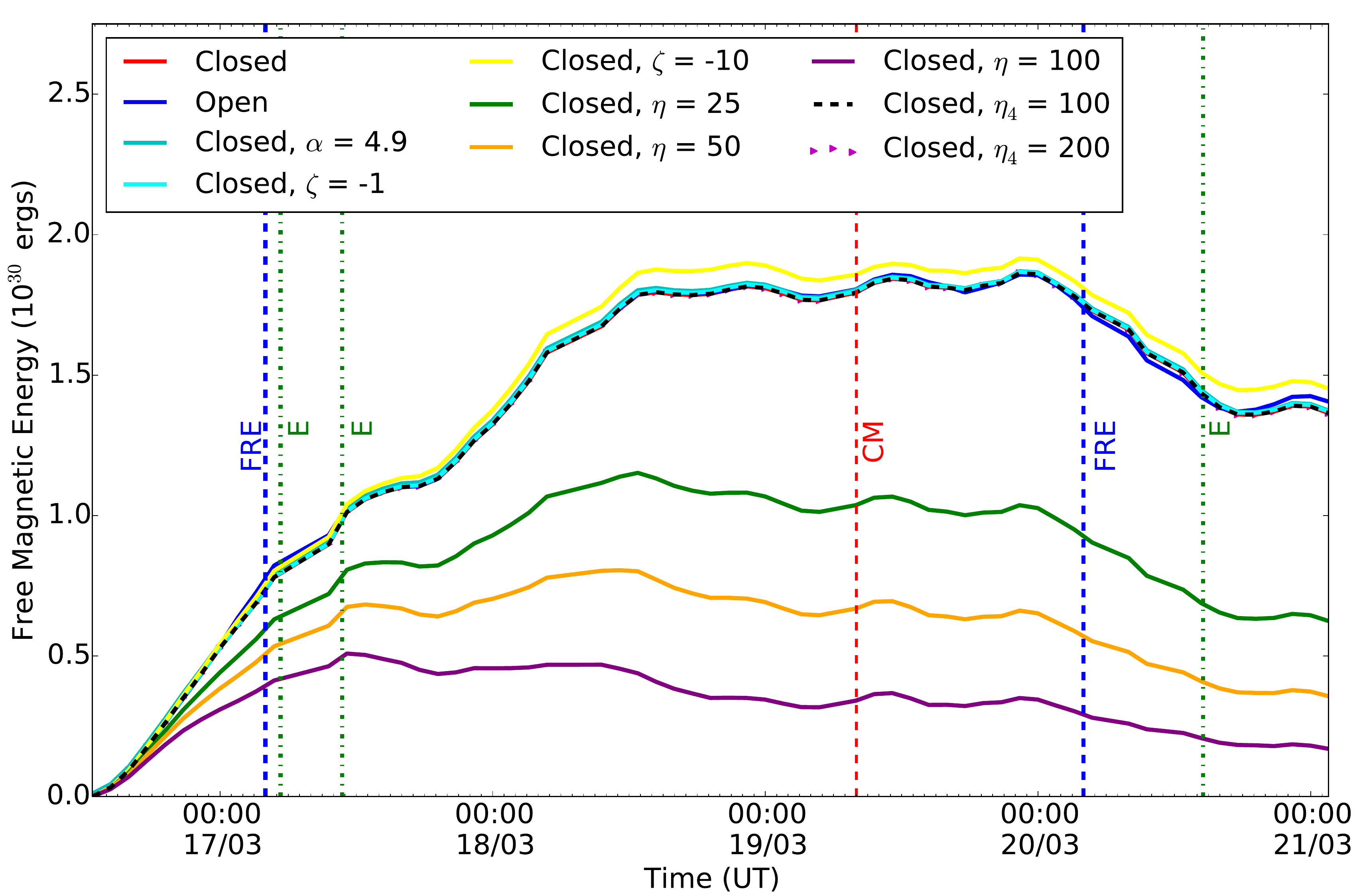}
\caption{The evolution of the free magnetic energy as a function of time for each simulation in Table \ref{tab:table1} calculated between 2012 March 16 12:46 UT and 2012 March 21 01:34 UT. The red dashed line shows the central meridian (CM) passage of AR 11437. The blue dashed and green dot dashed lines represent the times of the simulated flux rope eruptions (FRE) and of the observed eruptions (E), respectively. \label{fig10}}
\end{figure}

We also investigate the evolution of the free magnetic energy $E$, which is given by

\begin{equation}
E = \frac{1}{8 \pi} \int (\mathbf{B}^{2} - \mathbf{B}_{p}^{2}) d\tau,
\end{equation}

where $\mathbf{B}$ is the magnetic field of the NLFF field simulation and $\mathbf{B}_{p}$ is the potential magnetic field that is extrapolated from the same boundary conditions as the simulated coronal field \citep{Mackay-2011}. The evolution of the free magnetic energy with time for each simulation is shown in Figure \ref{fig10}.

The general evolution of the free magnetic energy will now be discussed. A similar trend in the free magnetic energy is seen for all ten simulations, apart from the three simulations that include the Ohmic diffusion term, where the free magnetic energy is systematically lower.
Initially, there is an increase of free magnetic energy corresponding to the emergence of flux and the counter-clockwise rotation of the bipole. The first simulated flux rope eruption occurs during this period. On 2012 March 18 at 11:12 UT the free magnetic energy reaches a maximum value of $\sim$ 1.8 $\times$ 10$^{30}$ ergs and remains roughly constant until 2012 March 20. This mostly corresponds to the decay phase of the AR when flux cancellation is taking place and the AR rotates clockwise. The free magnetic energy then starts to decrease and the eruption of the second simulated flux rope occurs. At the end of the evolution a small increase in magnetic flux occurs due to the emergence of a small bipole and the AR rotates counter-clockwise. This results in the small and final increase in the free magnetic energy.

\begin{figure}
\plotone{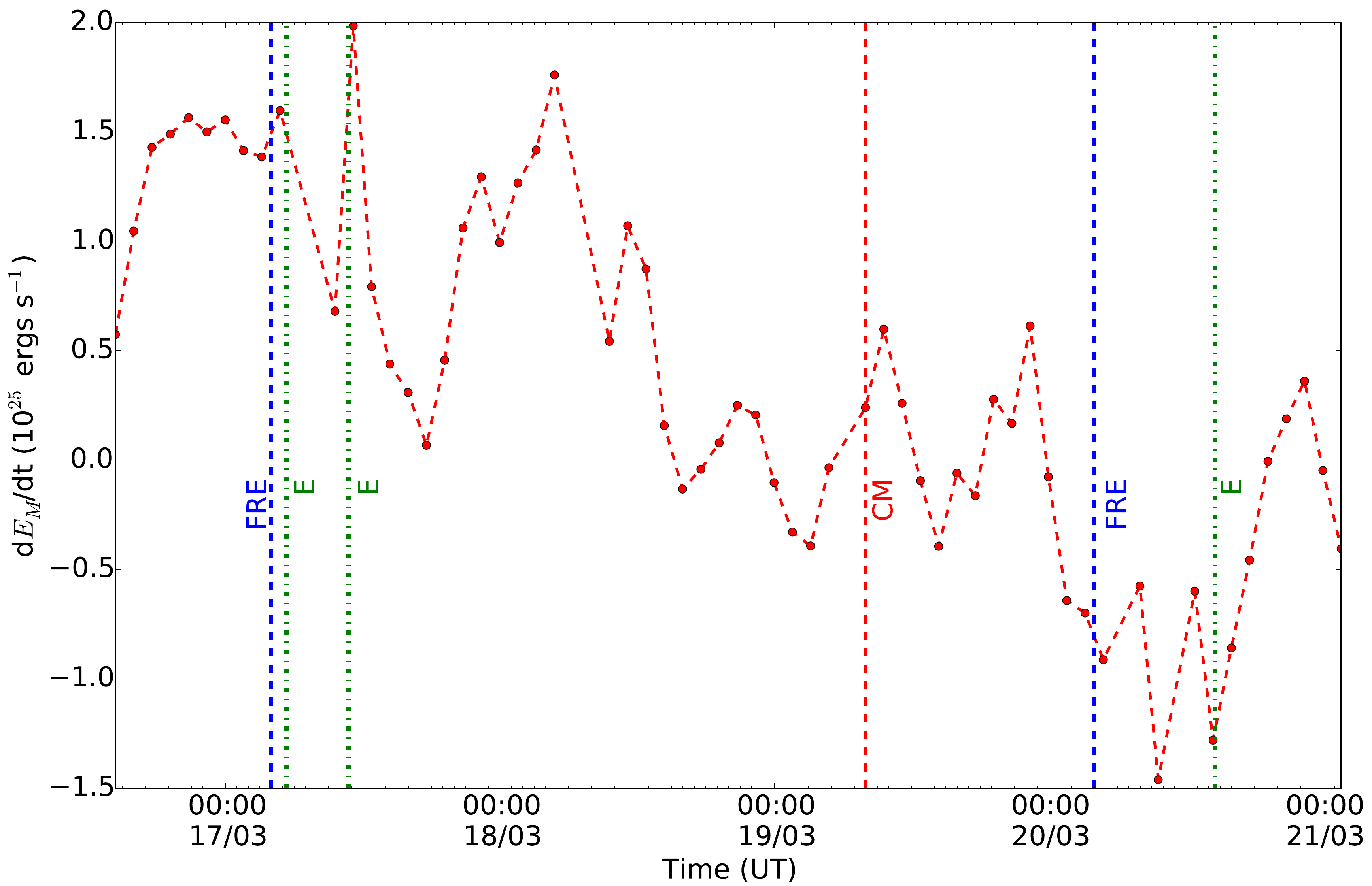}
\caption{The rate of change of free magnetic energy as a function of time for closed boundary conditions (simulation 1 in Table \ref{tab:table1}) between 2012 March 16 12:46 UT and 2012 March 21 01:34 UT. The red dashed line indicates when AR 11437 crosses central meridian (CM). The blue dashed and green dot dashed lines represent the timings of the simulated flux rope eruptions (FRE) and the eruptions in the observations (E), respectively. \label{fig11}}
\end{figure}

The timings of the simulated flux rope eruptions coincide with changes in the evolution of the free magnetic energy. Around the time of the first simulated flux rope eruption the free magnetic energy is increasing. The rate at which the free magnetic energy increases slows after the first eruption has occurred. Similarly, the free magnetic energy starts to decrease prior to the eruption of the second simulated flux rope. This variation is more apparent in the rate of change of free magnetic energy, which is shown in Figure \ref{fig11}. The rate of change of the free magnetic energy decreases by 9.2 $\times$ 10$^{24}$ erg s$^{-1}$ and 1.5 $\times$ 10$^{25}$ ergs s$^{-1}$ around the times of the first and second simulated flux rope eruptions, respectively. The decrease in the rate of change of free magnetic energy around the timings of the eruption is small due to the evolution of the applied boundary and the continual injection of a Poynting flux at the base of the computational box.


\subsubsection{Relative Helicity}

\begin{figure}
\plotone{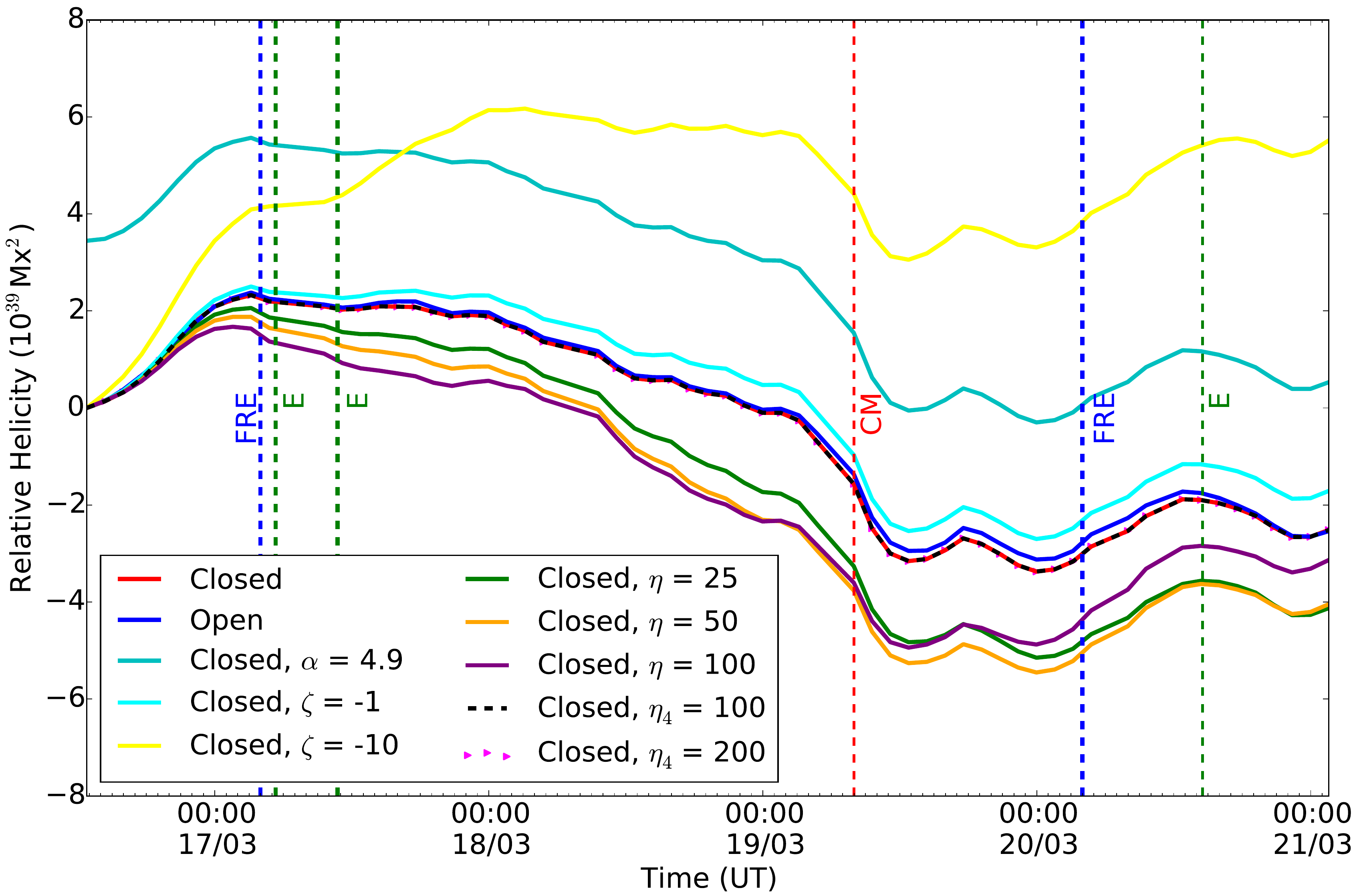}
\caption{Relative helicity evolution as a function of time between 2012 March 16 12:46 UT and 2012 March 21 01:34 UT. The red dashed line indicates when AR 11437 crosses central meridian (CM). The blue dashed and green dot dashed lines represent the timings of the simulated flux rope eruptions (FRE) and the eruptions in the observations (E), respectively. \label{fig12}}
\end{figure}

Small-scale random motions at the photospheric surface inject magnetic helicity along with free magnetic energy into the coronal magnetic field. Magnetic helicity is a topological measure of the twist and linkage of magnetic field lines and is approximately conserved during the process of magnetic reconnection \citep{Berger-1999}. To study the injection and evolution of helicity in the simulations the relative helicity $H_{R}$ is calculated as follows

\begin{equation}
H_{R} = \int (\mathbf{A} \cdot \mathbf{B}) d\tau - \int (\mathbf{A_{p}} \cdot \mathbf{B_{p}}) d\tau,
\end{equation}

where $\mathbf{A}$ is the magnetic vector potential and $\mathbf{B}$ is the magnetic flux density of the simulation (see \citealt{Mackay-2011}). The vector potential and magnetic flux density of the potential field with the same normal field component and boundary conditions is given by $\mathbf{A_{p}}$ and $\mathbf{B_{p}}$, respectively. The evolution of the relative helicity of the simulations is shown in Figure \ref{fig12}.

The relative helicity in each simulation shows a similar evolutionary behaviour throughout the time period studied. At the beginning of the evolution the relative helicity for the majority of the simulations initially increases to $\sim 2.3 \times 10^{39}$ Mx$^{2}$ as positive helicity is injected into the corona. For the case where the force-free parameter $\alpha = 4.9 \times 10^{-8}$ m$^{-1}$ the initial relative helicity is  $\sim 3.5 \times 10^{39}$ Mx$^{2}$ due to the LFF field initial condition. While the value is initially higher it follows the same trend as the other simulations. By introducing an additional injection of helicity, particularly in the case $\zeta = 10$ km$^{2}$ s$^{-1}$, the helicity increases as expected. For the majority of the simulations the helicity is positive during the first half of the evolution, which follows the hemispheric rule of positive helicity being dominant in the southern hemisphere \cite{Pevtsov-1995}. However, once the active region crosses central meridian the helicity sign is negative. The evolution of helicity is very similar to that of the tilt angle suggesting that the dominant source of helicity injection is the large scale rotation of the active region.

\section{Summary \& Discussion} \label{sec:con}

In this study, we have simulated the coronal evolution of AR 11437 during its entire lifetime, from emergence to decay (2012 March 16 12:46 UT to 2012 March 21 01:34 UT). The coronal field has been simulated using the magnetofrictional relaxation technique of \citet{Mackay-2011} with \textit{SDO}/HMI LoS magnetograms as lower boundary conditions. By applying this method it was possible to replicate the main coronal features and evolution of the active region. Observations from \textit{SDO}/AIA show that the coronal field of the active region becomes sheared and three eruptions occur, two take place in the emergence phase and one in the decay phase. The simulation was able to reproduce the sheared coronal structure and two out of the three observed eruptions. We also conducted a parameter study to include global parameters such as Ohmic diffusion, hyperdiffusion and helicity injection in the simulation. The addition of these terms in the simulations does not effect the overall coronal evolution. Therefore, the magnetofrictional relaxation technique of \citet{Mackay-2011} using LoS magnetograms as lower boundary conditions is very effective in simulating the coronal evolution of AR 11437 and its eruptions.

The technique was particularly successful in capturing the formation and evolution of a sheared coronal structure present in the observations. In the simulation a flux rope forms in the north of the active region at the same location as the sheared coronal structure. In the early stages of the coronal evolution there are some discrepancies between the simulated field and coronal observations in the south of the AR. The simulated flux rope structure does not extend as far south as the coronal loops. The deviation between the observations and simulation decreases with time and both are in good agreement during the later stages of evolution. This variation is most likely caused by the potential field initial condition of the simulation. These findings are very similar to that of \citet{Gibb-2014}, where the simulated field captures the formation and evolution of an active region sigmoid. 

During the emergence phase, AR 11437 rotates counter-clockwise and a dark loop system that is located to the west of the active region erupts on 2012 March 17 at 05:09 UT. The simulated flux rope at this location starts to rise and is found to erupt roughly 1 hour before the ejection of the dark loop system in the observations. As the dark loop system is located directly next to the flux rope in the simulation it could therefore be a part of this pre-eruptive structure. Signatures of eruption include the formation of kinked field lines and post-reconnection loops below the simulated flux rope as it rises. The evolution of the coronal field is driven with LoS magnetograms that have a cadence of 96 minutes and therefore, the eruption occurs within the time resolution of the data. In the observations there are no white-light signatures associated with this eruption and when the simulated flux rope erupts it does not rise to the upper section nor leave the computational box even when open top boundary conditions are applied. In the AIA 171 \AA\ observations there is an expansion of coronal loops and a small brightening. This suggests that the observed eruption could be confined or that only a partial eruption of the sheared structure occurs. A filament then forms and erupts on 2012 March 17 at 10:45 UT. The second eruption is not captured in the simulations as it occurs only 3 time steps (6 hours) after the previous eruption making it hard to distinguish.


As the active region enters its decay phase flux cancellation occurs at the internal PIL. Approximately 1.3 $\times 10^{20}$ Mx of magnetic flux cancels during a time period of 1.5 days. This amounts to 27~\% of the peak active region flux being cancelled. The amount of flux cancelled is very similar to that found in \citet{Yardley-2017} despite the difference in methods used to calculate the magnetic flux. This is also consistent with previous studies such as \citep{Green-2011, Baker-2012, Yardley-2016}. The active region also rotates clockwise during this period. In the simulations the flux rope, which formed along the internal PIL, has expanded and grown in length, resulting in a larger structure that is consistent with the J-shaped structure in the coronal observations. The expansion and growth of the simulated flux rope provides additional evidence that the first eruption was likely to be confined. The large flux rope structure forms during the decay phase of the active region evolution suggesting that the process of flux cancellation and associated reconnection at the internal PIL was responsible. Hence, the evolutionary sequence of this structure is consistent with the model of \citet{vB-1989}. The final eruption of the simulated flux rope structure occurs roughly 10 hours before the eruption of the observed J-shaped coronal structure. Furthermore, the flux rope rises once again but does not leave the computational box.

When varying the boundary and initial conditions and including additional parameters such as Ohmic diffusion, hyperdiffusion and helicity injection in the simulations, the evolution of the coronal field does not change considerably. This suggests that a key element in reproducing the main evolutionary features of the active region is using the normal component of the magnetic field from LoS magnetograms to drive the simulations. However, when including Ohmic diffusion the simulated field has less twist than the other cases. This is expected as Ohmic diffusion acts to decrease poloidal flux in the simulation and therefore decreases the twist of the simulated field.

For the majority of the simulations conducted the flux rope erupts roughly 1 and 10 hours before the first and third eruptions in the observations, respectively. The inclusion and increased values of the Ohmic diffusion term causes the timings of the eruptions of the simulated flux rope to differ to the other cases. When Ohmic diffusion takes values of 25, 50 and 100 km$^{2}$ s$^{-1}$ the simulated flux rope erupts roughly 2, 5 and 7 hours after the first observed eruption, respectively. This is due to the fact that the flux rope structure is less twisted than in the other cases. For the final eruption, the simulated flux rope erupts roughly 13 and 21 hours before the observed eruption for $\eta$ values of 50 and 100 km$^{2}$ s$^{-1}$. At this point in the simulation the flux rope twist is still small but the rope is more inflated compared to the other simulations. The early eruption of the flux rope is due to the eruption taking place during the later stages of the simulation. This allows a continual build-up of stress to occur over a longer period as the simulation fails to capture the second observed eruption. When Ohmic diffusion is included the overlying field remains more potential, the restoring forces acting on the flux rope are weaker, and as a result, the eruption occurs even earlier in the simulation.


The free magnetic energy was calculated as a function of time and in all ten simulations followed the same evolutionary trend. Although, the values of free magnetic energy were systematically lower for the simulations that included Ohmic diffusion. Initially, the free magnetic energy increases due to flux emergence. The rate of change of free magnetic energy is seen to decrease by 9.2 $\times$ 10$^{24}$ erg s$^{-1}$ but remains positive after the first simulated eruption. The free magnetic energy continues to increase after the emergence phase. This is due to the injection of energy through small-scale convective motions which inject a Poynting flux into the corona \citep{Mackay-2011}. Finally, during the final eruption the rate of change of free magnetic energy decreases by -1.5 $\times$ 10$^{25}$ erg s$^{-1}$ as the energy overall decreases. 

We also study the injection and evolution of the relative helicity with time for each simulation. As for the free magnetic energy, the relative helicity shows the same trend in all ten simulations apart from in two cases that have consistently higher values of helicity. Firstly, in the case where a LFF field initial condition is used and the force-free parameter $\alpha$ is $4.9 \times 10^{-8}$ m$^{-1}$. Secondly, when there is a large helicity injection of $10$ km$^{2}$ s$^{-1}$. The helicity is initially increasing and positive, which agrees with the hemispheric rule that states positive helicity dominates in the southern hemisphere \citep{Pevtsov-1995}. The helicity then decreases and as the active region crosses central meridian becomes negative. This is because the active region rotates clockwise and the evolution of helicity follows the same trend as the evolution of the active region tilt angle. This suggests that active region rotation is a large source of helicity injection in the corona \citep{Gibb-2014}.

In this study, we have used the magnetofrictional technique of \citet{Mackay-2011} to successfully simulate the coronal evolution of AR 11437 along with two out of three observed eruptions from the active region. This technique uses a time series of \textit{SDO}/HMI LoS magnetograms as lower boundary conditions. When including additional global parameters such as Ohmic diffusion, hyperdiffusion and an injection of helicity the overall coronal evolution of the simulated field and eruption times did not change significantly. This shows that the technique of \citet{Mackay-2011} only requires the normal component of the magnetic field from LoS magnetograms as lower boundary conditions to successfully reproduce the coronal evolution and the build-up to the eruption of an active region.

\acknowledgements

The authors would like to thank \textit{SDO}/HMI and AIA consortia for the data, and also being able to browse data through JHelioviewer (http://jhelioviewer.org/). This research has made use of SunPy, an open-source and free community-developed solar data analysis package written in Python \citep{SunPy-2015}. S.L.Y. would like to acknowledge STFC for support via the Consolidated Grant SMC1/YST025. D.H.M. would like to thank STFC and the Levehulme Trust for financial support. L.M.G. is grateful to the Royal Society for a University Research Fellowship and the Leverhulme Trust.

\appendix

\section{Magnetogram Clean-up Process} \label{sec:A}

This study used magnetograms taken from the 720s data series made available by the Helioseismic and Magnetic Imager (HMI) on board the \textit{Solar Dynamics Observatory (SDO)}. The full-disk LoS magnetograms are computed from filtergrams, which are recorded by the vector camera, have a pixel size of 0.5" and a noise level of 10 G. In total, 62 magnetograms are used with a 96 minute cadence during the time period beginning 2012 March 16 12:46 UT until 2012 March 21 01:34 UT. A cosine correction is applied to the magnetic field values to estimate the radial component of the magnetic field. Each magnetogram, which contains the corrected radialised field values, is de-rotated to account for area foreshortening effects at large centre-to-limb angles. Cut-out images are taken from the corrected and de-rotated magnetograms centred on the AR with a size of 278 $\times$ 279
pixels.

The cleaning procedure that will now be described is very similar to that in \citet{Gibb-2014}. It is apparant in Figures \ref{fig1} and \ref{fig3} that the noise level is high in the raw magnetograms. This is particularly evident in the magnetograms taken during the early and late phase of the active region evolution when the distance from central meridian is large. To be able to use the magnetograms as boundary conditions in the simulation several clean-up procedures were applied. Firstly, time-averaging is applied to the magnetograms using a Gaussian kernel as follows

\begin{equation}
C_{i} = \frac{ \sum^{62}_{j=1}  \mathrm{exp}(-[i-j]/\tau)^{2} F_{j} } {\sum^{62}_{j=1} \mathrm{exp}(-[i-j]/\tau)^{2} },
\end{equation}

where $C_{i}$ represents the $i$th cleaned frame and ranges from 1 to 62, $F_{j}$ is the $j$th raw frame, and $\tau$ is the separation of frames where the weighting falls by $1/e$. In this case we set the separation to be two frames so $\tau$ = 2. Therefore, each of the cleaned frames is a linear combination of the 62 raw frames however, the two frames previous and after the current frame have the highest weighting. This procedure is used to retain large scale active region features and remove random noise.

\begin{figure}[h]
\plotone{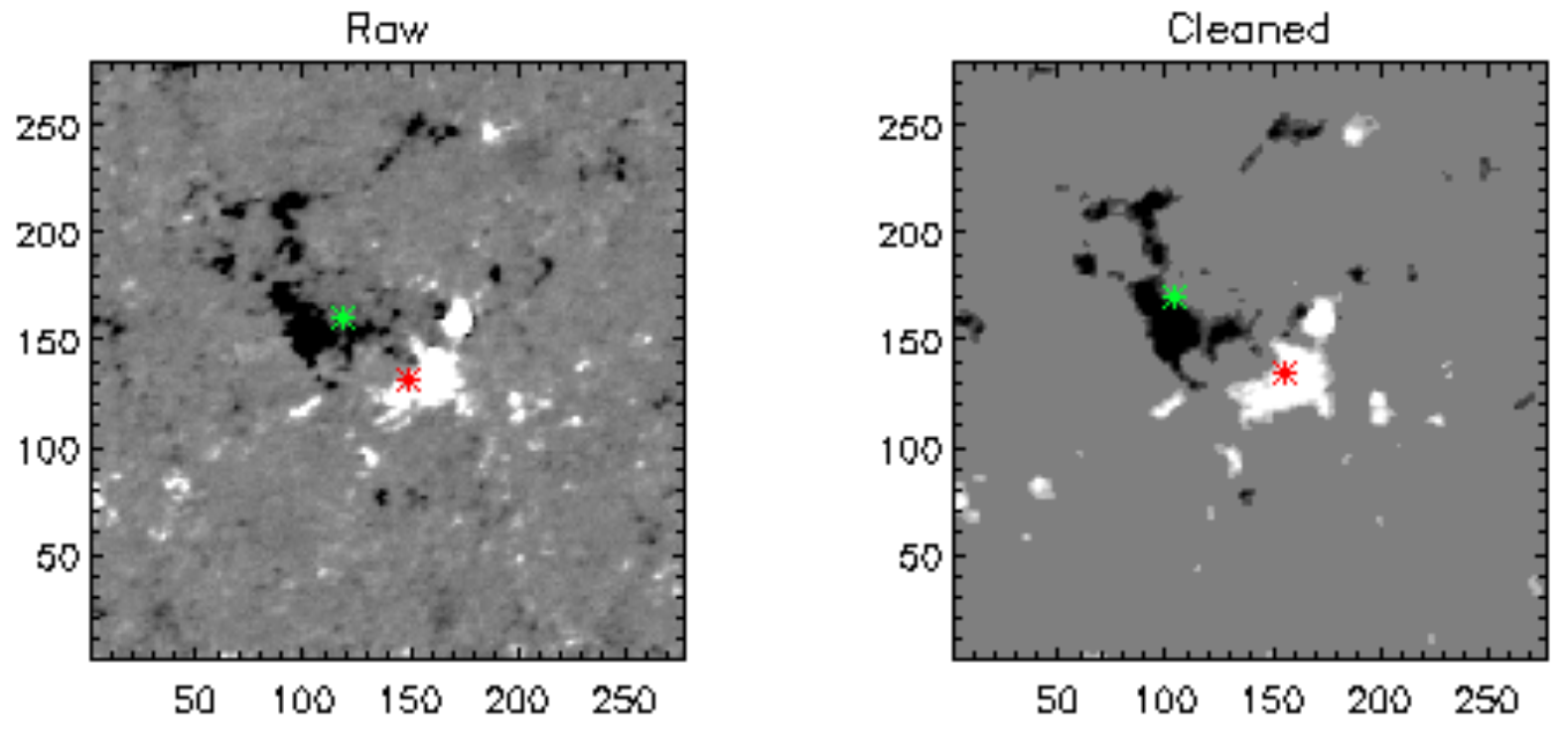}
\caption{The raw and cleaned magnetograms for frame 15 of the simulation taken on 2012 March 17 at 15:59 UT when AR 11437 is at peak unsigned magnetic flux. The saturation levels of the magnetic field for both magnetograms is set to $\pm$100 G. The flux-weighted central coordinates for the positive (negative) polarity is represented by the red (green) asterisk. \label{fig13}}
\end{figure}

In this study we are only interested in analysing the evolution of the large scale magnetic field of the active region and not small-scale features such as network elements or quiet Sun magnetic field. Therefore, small-scale isolated magnetic features are removed pixel-by-pixel by considering the eight nearest neighbours of each pixel. If less than four of the nearest neighbours had the same sign of magnetic flux then the flux value of that pixel would be set to zero. This means that as the pixels at the edges of the magnetograms had less than eight neighbouring pixels their values were also set to zero. Additionally, the pixels that had magnetic flux values that were below the threshold of 25 Mx cm$^{-2}$ were assigned a value of zero as these pixels form part of the quiet Sun background magnetic field.

The final clean-up procedure was applied to ensure that the magnetograms were flux balanced. This procedure is required if the top boundary conditions of the simulation are closed. The magnetograms were flux balanced by calculating the signed magnetic flux of each frame. For each frame the pixels of non-zero value were counted and the signed magnetic flux was divided by this total. The imbalanced magnetic flux per non-zero valued pixel was subtracted from every pixel of non-zero value. No pixels changed sign during the procedure of flux balancing as the maximum correction was less than the threshold of 25 Mx cm$^{-2}$. This was the threshold used to set the value of pixels that formed part of the background field to zero.

Figure \ref{fig13} shows the raw and cleaned magnetograms taken at the time of the peak unsigned magnetic flux of AR 11437 (2012 March 17 at 15:59 UT). The figure shows that the large scale magnetic features of the active region in the raw magnetogram are still present in the cleaned magnetogram but the small-scale magnetic features and noise has been removed. 

Several clean-up procedures have been applied including time-averaging, isolated feature removal, low flux value removal, and flux balancing to produce a series of cleaned magnetograms that show a smooth, continuous photospheric field evolution. Using these magnetograms as photospheric boundary conditions makes it easier to simulate the evolution of the coronal magnetic field as random noise and small-scale magnetic features can cause numerical problems in the magnetofrictional simulation.

\section{Tilt Angle Calculation} \label{sec:B}

To determine the tilt angle of an AR the centers of flux for the positive and negative polarity must be calculated as a function of time using

\begin{equation}
\mathbf{r_{\pm}} = \left( \int \mathbf{r} B_{\pm} dA \right) \left(\int B_{\pm} dA \right),
\end{equation}

where $\mathbf{r}$ is a position vector, $\pm$ represents the positive and negative polarity flux, and A is the area. The angle $\theta$ can then be calculated by

\begin{equation}
\theta = \arctan \left( \frac{(\mathbf{r_{-}} - \mathbf{r_{+}})_{y}} {( \mathbf{r_{-}} - \mathbf{r_{+}})_{x}} \right),
\end{equation}

where the $x$, $y$ subscripts are the x and y components of $\mathbf{r_{\pm}}$, respectively. This follows the method outlined in \citet{Gibb-2014} however, we take the tilt angle to be $180 - \theta$ in this case following a similar definition to \citet{Tian-2001}.

\end{document}